\documentclass[11pt,here,twoside,a4paper]{report}

\usepackage{a4}
\usepackage{amstext}
\usepackage{latexsym}
\usepackage{amsmath}
\usepackage{tabularx}
\usepackage{array}
\usepackage{amssymb}
\usepackage{multirow}
\usepackage{psboxit}
\usepackage{epsfig}
\usepackage{graphics}
\usepackage{here}
\usepackage{amsfonts}

\newcommand{\clearemptydoublepage}{\newpage{\pagestyle{empty}\cleardoublepage}}

\begin{document}


\thispagestyle{empty}
\textheight 22cm
\begin{titlepage}

 $ $
\vspace{3cm}

\begin{center}
{\Huge \bf Duality for the Non-Specialist}
\end{center}
\vspace{1cm}

\begin{center}
{\large {\bf Svend E. Hjelmeland}\\ ~\\
		\it Institute of Physics, University of Oslo\\
		\it P.O. BOX 1048, N-0316 Blindern, Oslo 3,\\
		\it  NORWAY}
\end{center}
\vspace{0.25cm}
\begin{center}
{\large {\bf Ulf Lindstr\"{o}m}\\ ~\\
                \large \it ITP, Stockholm University\\
		\it Box 6730, Vanadisv\"{a}gen 9, S-113 85, Stockholm,\\
		\it  SWEDEN}
\end{center}
\begin{center}
{\large and}
\end{center}
\begin{center}
		{\it Institute of Physics, University of Oslo\\
		\it P.O. BOX 1048, N-0316 Blindern, Oslo 3,\\
		\it NORWAY}
\end{center}
\end{titlepage}
\clearemptydoublepage
\textheight 22cm
\begin{titlepage}

 $ $
\vspace{3cm}

\begin{center}
{\Huge \bf Duality for the Non-Specialist}
\end{center}
\vspace{1cm}

\begin{center}
{\large {\bf Svend E. Hjelmeland\footnote{e-mail address: s.e.hjelmeland@fys.uio.no}}\\ ~\\
		\it Institute of Physics, University of Oslo\\
		\it P.O. BOX 1048, N-0316 Blindern, Oslo 3,\\
		\it  NORWAY}
\end{center}
\vspace{0.25cm}
\begin{center}
{\large {\bf Ulf Lindstr\"{o}m\footnote{e-mail address: ul@vanosf.physto.se, ulfl@boson.uio.no}}\\ ~\\
                \large \it ITP, Stockholm University\\
		\it Box 6730, Vanadisv\"{a}gen 9, S-113 85, Stockholm,\\
		\it  SWEDEN}
\end{center}
\begin{center}
{\large and}
\end{center}
\begin{center}
		{\it Institute of Physics, University of Oslo\\
		\it P.O. BOX 1048, N-0316 Blindern, Oslo 3,\\
		\it NORWAY}
\end{center}
\vspace{1cm}
\begin{center}
{\bf Abstract}
\end{center}
{This is a review of the basics of duality as applied to $p$-forms and $\sigma$-models. The ideas are introduced by way of worked examples, often quite detailed. Our approach is very pedestrian and the presentation is aimed at non-specialists, such as e.g. graduate students.}

\end{titlepage}
\clearemptydoublepage
\chapter*{~Preface}
\thispagestyle{empty}

The report you hold in your hands represents the written version of a series of lectures that were presented by U.L. at the University of Oslo in the fall of 1996. The notes were checked, elaborated on and texed by S.E.H. We took the opportunity to add some useful references (with no attempt what so ever at completeness). To further improve the notes we also added a section on Poisson-Lie $T$-duality that was not included in the original lectures.
\\
\\

\hspace*{9cm}Oslo, March 1997\\ \\
\hspace*{10cm}S.E.Hjelmeland\\
\hspace*{10cm}U.Lindstr\"{o}m

\newpage
\newpage				
\clearemptydoublepage

\pagenumbering{roman}

\thispagestyle{plain}
\tableofcontents
\clearemptydoublepage
\vfill\pagebreak


\chapter*{~Introduction}
\addcontentsline{toc}{chapter}{Introduction}

Dualization has by know appeared in several different contexts in theoretical physics. A few of those are: Kramers-Wannier duality, dual models, Hodge dual, dual maps, scalar-tensor duality, electric-magnetic duality, Montonen-Olive duality, the low energy effective action duality constructed by Seiberg, and the recent string dualities ($S$-duality, $T$-duality, $U$-duality). In this lecture some of these dualities will be outlined in more detail.

{\it Here duality will mean that there exist two equivalent descriptions of a model using different fields}. A classical example is the {\it scalar-tensor duality} in 4D. A free Klein-Gordon field $\phi$ has an equivalent description in terms of a free antisymmetric tensor field $A_{\mu\nu}$\footnote{We will see that this as well as electric-magnetic duality are special cases of p-form duality in $D$-dimensions where a $p$-form is dual to a $D-(p+2)$-form.}. The relation between the fields is describable as a {\it Legendre transform}, but an explicit description of one as a function of the other would be non-local and non-linear. Only in certain 2D-dualities do we have an explicit relation, such as in the case of the duality between the {\it massive Thirring model} $S(\psi)$ and the {\it Sine-Gordon model} $S(\phi)$ where $\phi\sim\bar{\psi}\psi$, is a bound state from the point of view of the Thirring model \cite{kn:RAJ}. The reason why such cases are important and interesting is the fact that {\it duality typically exchanges the coupling regimes}: $g\rightarrow 1/g$, then the weak coupling regime in one model is the strong coupling regime in the other and vice versa. Knowing the explicit relation thus allows perturbative calculations in the variables of the original theory both in the strong and weak coupling regimes.

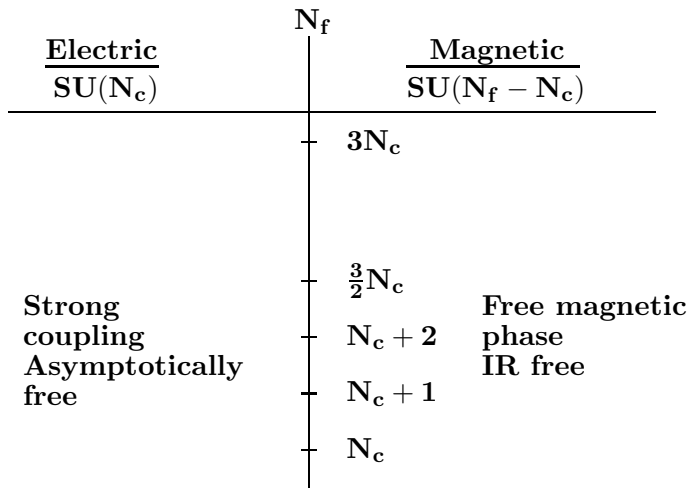
\begin{figure}[t]  
\begin{center}
\setlength{\unitlength}{1mm}
\begin{picture}(100,60)
\put(48,61){$\bf{N_{f}}$}
\put(10,50){\line(1,0){86}}
\put(50,0){\line(0,1){60}}
\put(15,57){\bf{Electric}}
\put(66,57){\bf{Magnetic}}
\put(15,56){\line(1,0){15}}\put(63,56){\line(1,0){23}}
\put(16,52){$\bf{SU(N_{c})}$}\put(63,52){$\bf{SU(N_{f}-N_{c})}$}
\put(49,46){\line(1,0){2}}\put(55,45){$\bf{3N_{c}}$}
\put(49,5){\line(1,0){2}}\put(55,4){$\bf{N_{c}}$}
\put(49,12.5){\line(1,0){2}}\put(55,11.5){$\bf{N_{c}+1}$}
\put(49,20){\line(1,0){2}}\put(55,19){$\bf{N_{c}+2}$}
\put(49,27.5){\line(1,0){2}}\put(55,26.5){$\bf{\frac{3}{2}N_{c}}$}
\put(12,23){\bf{Strong}}\put(12,19){\bf{coupling}}\put(12,15){\bf{Asymptotically}}
\put(12,11){\bf{free}}
\put(73,23){\bf{Free magnetic}}\put(73,19){\bf{phase}}\put(73,15){\bf{IR free}}
\end{picture}
\caption{\small For $N_{c}+1\leq N_{f}\leq\frac{3}{2}N_{c}$ the low energy description of the models on the left and right hand side are dual.}
\label{low-energ}
\end{center}
\end{figure}

Imagine for a moment that QCD had a dual description and that we knew the explicit transformations. We would then have perturbative control over both the asymptotically free and the confined phase. This of course is too much to hope for, but recent years have seen a remarkable development in field theory along these lines. Namely, Seiberg and Witten \cite{kn:SW1} have solved the $N=2$ supersymmetric Yang-Mills theory in a way that utilizes duality - an ``electric-magnetic'' duality of the kind conjectured by Montonen and Olive \cite{kn:MO} and known for $N=4$ - and, they also showed that breaking $N=2$ down to $N=1$ gives {\it electric confinement}. So, there exists a semi-realistic theory with some of the desired properties. Further, through the work of Seiberg and collaborators \cite{kn:SEI} a {\it ``low energy effective duality''} has been shown to exist in certain $N=1$ supersymmetric field theories, with colour gauge groups $SU(N_{c})$ and flavour group $SU(N_{f})$ (see fig.\ref{low-energ} on the next page).

Now after this discovery by Seiberg in field theory it still remained an open question whether similar relations exist in string theory. Field theory being the low energy limit of strings it is clear that string duality implies field theory duality, but not the other way around. Dualities in string theory have been
conjectured for a long time but only recently has there been sufficient evidence to believe the duality to exist. In fact, string theory is presently undergoing a very rapid development, (``2nd revolution''), which is based on the discovery of D-branes\footnote{D-branes are solitonic solutions to the string equations that are ``branes'' where open strings end (i.e. where those strings have Dirichlet boundary conditions).}, the existence of dualities, etc. \cite{kn:POL}. Field theory duality multiplets contain elementary quanta and smooth classical configurations (magnetic monopoles). String duality multiplets contain these (elementary quanta, strings) plus singular configurations (black holes) and D-branes.
\begin{figure}[h]
\centering
\setlength{\unitlength}{1mm}
\begin{picture}(90,60)
\put(0,0){\mbox{\epsfig{file=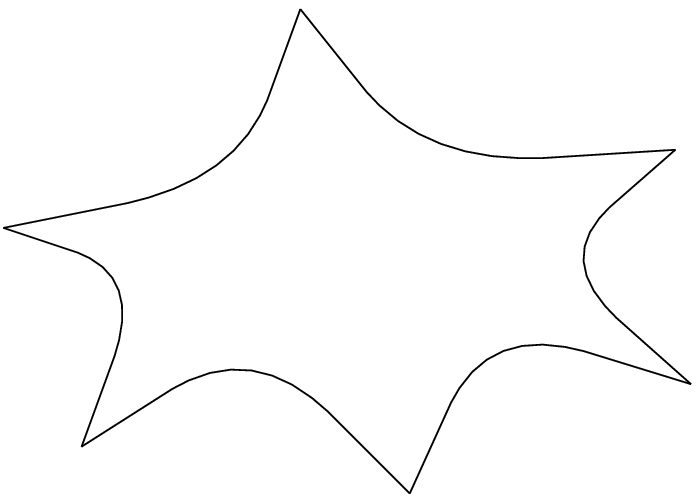,height=6cm}}}
\put(-18,33){\bf{Type IIA}}\put(-9,6){\bf{Type IIB}}\put(52,0){\bf{Type I}}
\put(39,59){\bf{M-theory}}\put(85,43){\bf{SO(32)}}\put(85,39){\bf{heterotic}}
\put(86,15){${\bf{E_{8}}}{\bf{\times}}{\bf{E_{8}}}$}\put(86,11){\bf{heterotic}}
\end{picture}
\caption{\small The moduli space of string vacua.}
\label{moduli}
\end{figure}
\newpage

String dualities were discovered in investigation of dualities that relate various weakly coupled string theories. In fact the picture that now emerge in string theory may be depicted something like that of Fig.\ref{moduli} (borrowed from \cite{kn:POL}). The known models correspond to weak coupling and the full moduli space is largely unchartered. In particular one sees evidence of an extra eleventh dimension in one direction in moduli-space. The low-energy limit of this unknown {\it 11D M-Theory} is {\it 11D supergravity}.

After this detour into the unknown, let us return to basics!

\clearemptydoublepage 
\chapter{Scalar-Tensor Duality}
\pagenumbering{arabic}

\section{Field eqn's $\leftrightarrow$ Bianchi's}

Consider the actions for a massless free Klein-Gordon field\footnote{We will use a notation where the fields are independent unless they have an explicit argument ($F_{\mu}$ is and independent field while $F_{\mu}(\phi)$ is not.).} $\phi$ in $4D$
\begin{equation}
S_{\phi}=\frac{1}{2}\int d^{4}xF_{\mu}(\phi)F^{\mu}(\phi), \label{action-fi}
\end{equation}
where $F_{\mu}(\phi)\equiv\partial_{\mu}\phi$, and the action for a massless free anti-symmetric second rank tensor field $A_{\mu\nu}$ in 4D
\begin{equation}
S_{A}=\frac{1}{3!}\int d^{4}xF_{\mu\nu\rho}(A)F^{\mu\nu\rho}(A), \label{action-A}
\end{equation}
where $F_{\mu\nu\rho}(A)\equiv\partial_{[\mu}A_{\nu\rho]}$. 

The field equation and the Bianchi identities for the free Klein-Gordon field are
\begin{eqnarray}
\partial_{\mu}F^{\mu}(\phi)    
&=&0\quad\text{(field equation)} \\ \nonumber \\
{\partial}_{\mu}^{~~*}F^{\mu\nu\rho}(\phi)     
&=&0\quad\text{(Bianchi identities)}
\end{eqnarray}
and for the free anti-symmetric tensor field we find 
\begin{eqnarray}
{\partial}_{\mu}^{~~*}F^{\mu}(A)     
&=&0\quad\text{(Bianchi identity)} \label{field-eq1} \\ \nonumber \\
\partial_{\mu}F^{\mu\nu\rho}(A)    
&=&0\quad\text{(field equations),}
\end{eqnarray}
where ${~}^{*}F^{\mu\nu\rho}(\phi)\equiv\epsilon^{\mu\nu\rho\sigma}F_{\sigma}(\phi)$ and ${~}^{*}F^{\mu}(A)\equiv\frac{1}{3!}\epsilon^{\mu\nu\rho\sigma}F_{\mu\nu\rho}(A)$.

The key observation here is that the field equation for the free Klein-Gordon field looks like the Bianchi identity for the free anti-symmetric field, and vice versa. In fact there exists a framework where it is seen that the two theories represent the same physics. A change from one description to the other interchanges the role of field equations and Bianchi identities. Let us look into this in more detail. Consider the so called {\it parent action} of $S_{A}$
\begin{equation}
S_{F,\phi}=\int d^{4}x\left(aF_{\mu\nu\rho}F^{\mu\nu\rho}+b\phi\partial_{\mu}^{~*}F^{\mu}\right), \label{parent1}
\end{equation}
where the scalar field $\phi$ is a Lagrange multiplier and $F_{\mu\nu\rho}$ is an independent field ($F\neq dA$). Varying $S_{F,\phi}$ with respect to $\phi$ gives directly
\begin{equation}
\delta\phi :~~~~~\partial_{\mu}^{~~*}F^{\mu}=0=\frac{1}{3!}\epsilon^{\mu\nu\rho\sigma}\partial_{\mu}F_{\nu\rho\sigma}.
\end{equation}
Hence, we force the field $F_{\nu\rho\sigma}$ to satisfy the Bianchi identity (eq.(\ref{field-eq1})). Thus, we may write $F_{\nu\rho\sigma}=\partial_{[\nu}A_{\rho\sigma]}$. Plugging this back into the action (\ref{parent1}) and choosing $a=\frac{1}{3!}$ we recover the $S_{A}$ of (\ref{action-A}). We have thus shown that (\ref{parent1}) is (classically) equivalent to (\ref{action-A})\footnote{Proving equivalence by substituting the solution of one set of field equations into the parent action is slightly risky, and requires some care for certain cases. The safe approach is to compare the field equations.}. 

To show that (\ref{parent1}) is also equivalent to (\ref{action-fi}) we again consider $S_{F,\phi}$. With the above value of $a$ it reads
\begin{equation}
S_{F,\phi}=\frac{1}{3!}\int d^{4}x(F_{\mu\nu\rho}F^{\mu\nu\rho}+b\phi\epsilon^{\mu\nu\rho\sigma}\partial_{\mu}F_{\nu\rho\sigma}). \nonumber
\end{equation} 
Varying it now with respect to $F_{\mu\nu\rho}$ gives
\begin{equation}
\delta F_{\mu\nu\rho}:~~~~~F^{\mu\nu\rho}=-\frac{b}{2}\epsilon^{\mu\nu\rho\kappa}\partial_{\kappa}\phi .
\end{equation}
Putting this back into $S_{F,\phi}$ we obtain
\begin{eqnarray}
S_{F,\phi}&{\rightarrow}&\frac{1}{3!}\int d^{4}x\left(\frac{b^2}{4}\epsilon^{\lambda\mu\nu\rho}\epsilon_{\kappa\mu\nu\rho}\partial_{\lambda}\phi\partial^{\kappa}\phi+\frac{b^2}{2}\epsilon^{\mu\nu\rho\sigma}\epsilon_{\kappa\nu\rho\sigma}\phi\partial_{\mu}\partial^{\kappa}\phi\right)\nonumber \\
&=&\frac{1}{3!}\int d^{4}x\left(\frac{b^2}{4}(-3!)\delta^{\lambda}_{~\kappa}\partial_{\lambda}\phi\partial^{\kappa}\phi-\frac{b^2}{2}(-3!)\delta^{\mu}_{~\kappa}\partial_{\mu}\phi\partial^{\kappa}\phi\right) \nonumber \\
&=&\frac{b^2}{4}\int d^{4}x\partial_{\mu}\phi\partial^{\mu}\phi .
\end{eqnarray}
So, $S_{F,\phi}\rightarrow S_{\phi}$ if $b=\sqrt{2}$. 

Thus, from the parent action $S_{F,\phi}$ we have shown that $S_{\phi}$ and $S_{A}$ is dual to each other; the two actions represent the same physics (at least classically), but the physical description is given using different fields. The characteristic feature of this construction is that the field equations and the Bianchi identities are exchanged. This duality may be illustrated as in Fig.\ref{dual1-1} below.
\begin{figure}
\begin{center}
\setlength{\unitlength}{1mm}
\begin{picture}(100,70)
\put(46,60){$\bf{S}_{\bf{F},\bf{\phi}}$}
\put(46,57){\vector(-2,-3){23}}\put(54,57){\vector(2,-3){23}}
\put(22,40){$\bf{\delta F_{\mu\nu\rho}}$}\put(68,40){$\bf{\delta\phi}$}
\put(18,17){$\bf{S_{\bf{\phi}}}$}\put(78,17){$\bf{S_{A}}$}
\put(20,10){\vector(0,1){5}}\put(80,10){\vector(0,1){5}}
\put(20,10){\line(1,0){20}}\put(60,10){\line(1,0){20}}
\put(46,9){\bf{dual}}
\end{picture}
\caption{\small The parent action $S_{F,\phi}$ is (classically) equivalent to both $S_{\phi}$ and $S_{A}$ showing that $S_{\phi}$ and $S_{A}$ are dual to each other.}
\label{dual1-1}
\end{center}
\end{figure}
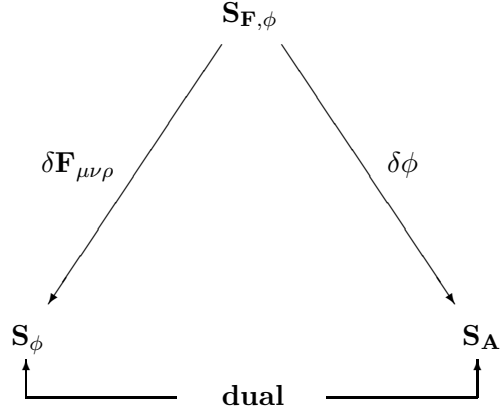

Parent actions are not unique. Another parent action which also shows that $S_{\phi}$ and $S_{A}$ are dual to each other is
\begin{equation}
S_{F,A}=\int d^{4}x\left(\tilde{a}F_{\mu}F^{\mu}+\tilde{b}A_{\mu\nu}\partial_{\rho}^{~~*}F^{\rho\mu\nu}\right).
 \end{equation}
Varying the action with respect to $A_{\mu\nu}$ we get right away
\begin{equation}
\delta A_{\mu\nu}:~~~~~\partial_{\rho}^{~~*}F^{\rho\mu\nu}=0=\epsilon^{\rho\mu\nu\sigma}\partial_{\rho}F_{\sigma},
\end{equation}
which is equivalent to $F_{\mu}=\partial_{\mu}\phi$. So $S_{F,A}\rightarrow S_{\phi}$ if $\tilde{a}=\frac{1}{2}$. If we instead vary the action with respect to $F_{\mu}$ we obtain
\begin{equation}
\delta F_{\mu}:~~~~~F^{\mu}=-\tilde{b}\epsilon^{\mu\nu\rho\sigma}\partial_{\nu}A_{\rho\sigma}.
\end{equation}
Putting the expression for $F^{\mu}$ back into $S_{F,A}$ one can easily verify that
\begin{equation}
S_{F,A}\rightarrow \frac{\tilde{b}^2}{3!2}\int d^{4}xF_{\mu\nu\rho}(A)F^{\mu\nu\rho}(A).
\end{equation}
So, if we put $\tilde{b}=\sqrt{2}$ we see that $S_{F,A}\rightarrow S_{A}$. The duality between $S_{\phi}$ and $S_{A}$ may be illustrated as shown in Fig.\ref{dual1-2}.
\begin{figure} 
\begin{center}
\setlength{\unitlength}{1mm}
\begin{picture}(100,70)
\put(46,60){$\bf{S}_{\bf{F},\bf{A}}$}
\put(46,57){\vector(-2,-3){23}}\put(54,57){\vector(2,-3){23}}
\put(25,40){$\bf{\delta A_{\mu\nu}}$}\put(68,40){$\bf{\delta F_{\mu}}$}
\put(18,17){$\bf{S_{\bf{\phi}}}$}\put(78,17){$\bf{S_{A}}$}
\put(20,10){\vector(0,1){5}}\put(80,10){\vector(0,1){5}}
\put(20,10){\line(1,0){20}}\put(60,10){\line(1,0){20}}
\put(46,9){\bf{dual}}
\end{picture}
\caption{\small The parent action $S_{F,A}$ is (classically) equivalent to both $S_{\phi}$ and $S_{A}$ showing that $S_{\phi}$ and $S_{A}$ are dual to each other.}
\label{dual1-2}
\end{center}
\end{figure}

\section{The shift property} 

Next, we want to discuss another construction of dual theories. Go back to (\ref{action-fi})
\begin{equation}
S_{\phi}=\frac{1}{2}\int d^{4}x\partial_{\mu}\phi\partial^{\mu}\phi. \nonumber
\end{equation}
This action has a {\it global symmetry}, namely it is invariant under a constant {\it shift} $\phi\rightarrow\phi +\epsilon$. We {\it gauge} this symmetry by introducing a field $V_{\mu}$;
\begin{equation}
\partial_{\mu}\phi\rightarrow D_{\mu}\phi\equiv\partial_{\mu}\phi +V_{\mu},
\end{equation}
and by letting
\begin{equation}
S_{\phi}\rightarrow S_{{\phi},V}\equiv\frac{1}{2}\int d^{4}xD_{\mu}\phi D^{\mu}\phi .
\end{equation}
Under a local shift, $(\epsilon =\epsilon(x))$, the action transforms into
\begin{eqnarray}
S_{{\phi},V}&=&S_{\phi}+\frac{1}{2}\int d^{4}x 2D_{\mu}\phi\delta(D^{\mu}\phi) \nonumber \\
&=&S_{\phi}+\int d^{4}xD_{\mu}\phi\left(\partial^{\mu}\epsilon+\delta V^{\mu}\right).
\end{eqnarray}
So, we have invariance if $\delta V^{\mu}=-\partial^{\mu}\epsilon$.

We add a term to the action $S_{\phi ,V}$ that ensures that $V^{\mu}$ is pure gauge
\begin{equation}
S_{W}=a\int d^{4}xA_{\mu\nu}^{~~~*}W^{\mu\nu},
\end{equation}
where ${~}^{*}W^{\mu\nu}\equiv\frac{1}{2}\epsilon^{\mu\nu\rho\sigma}W_{\rho\sigma}$ and $W_{\mu\nu}\equiv\partial_{[\mu}V_{\nu]}$. Varying $S_{W}$ with respect to $A_{\mu\nu}$, gives
\begin{equation}
\delta A_{\mu\nu}:~~~~~^{*}W_{\mu\nu}=0\Rightarrow V_{\mu}=\partial_{\mu}\lambda.
\end{equation}
One can show that $S_{\phi}$ is recovered if $\Box\lambda=0$ (modulo topological obstructions). This is also achieved if we choose a gauge $\lambda=0$, or redefine $\tilde{\phi}=\phi+\lambda$.

If we choose to integrate out $V_{\mu}$ instead, we find
\begin{equation}
D_{\mu}\phi=\frac{a}{2}\epsilon^{\rho\sigma\nu}_{~~~\mu}\partial_{\nu}A_{\rho\sigma}\equiv\frac{a}{3!2}\epsilon^{\rho\sigma\nu}_{~~~\mu}F_{\nu\rho\sigma}(A)
\end{equation}
so that
\begin{equation}
V_{\mu}=\frac{a}{3!2}\epsilon^{\rho\sigma\nu}_{~~~\mu}F_{\nu\rho\sigma}(A)-\partial_{\mu}\phi .
\end{equation}
Putting these expressions back into the action gives (temporarily suppressing the argument of $F$)
\begin{eqnarray}
S_{{\phi},V}+S_{W}&{\rightarrow}& \frac{1}{2}\int d^{4}x\frac{a^2}{(3!2)^{2}}\epsilon^{\rho\nu\sigma}_{~~~~\mu}\epsilon_{\kappa\tau\sigma}^{~~~~\mu}F_{\rho\nu\sigma}F^{\kappa\tau\lambda} \nonumber \\
& &+\frac{a}{2}\int d^{4}xA_{\mu\nu}\epsilon^{\mu\nu\rho\sigma}\partial_{\rho}\left(\frac{a}{3!2}\epsilon^{\alpha\beta\lambda}_{~~~~\sigma}F_{\lambda\alpha\beta}-\partial_{\sigma}\phi\right) \nonumber \\
&=&\frac{1}{2}\int d^{4}x\left\{\frac{a^2}{(3!2)^{2}}(-1!)\delta^{\rho}_{~[\kappa}\delta^{\nu}_{~\tau}\delta^{\sigma}_{~\lambda]}F_{\rho\nu\sigma}F^{\kappa\tau\lambda} \right.\nonumber \\
& &\qquad\qquad\left. -\frac{a^2}{3!2}\partial_{\alpha}A_{\mu\nu}(-1!)\delta^{\alpha}_{~[\lambda}\delta^{\mu}_{~\rho}\delta^{\nu}_{~\sigma]}F^{\lambda\rho\sigma} \right\} \nonumber \\
&=&\frac{a^2}{3!2}\int d^{4}xF_{\mu\nu\rho}(A)F^{\mu\nu\rho}(A).
\end{eqnarray}
Choosing $a=\sqrt{2}$ we see that $S_{\phi ,V}+S_{W}\rightarrow S_{A}$.

Conversely the action (eq.(\ref{action-A}))
\begin{equation}
S_{A}=\frac{1}{3!}\int d^{4}xF_{\mu\nu\rho}(A)F^{\mu\nu\rho}(A) \nonumber
\end{equation}
also has a symmetry under $A_{\mu\nu}\rightarrow A_{\mu\nu}+\epsilon_{\mu\nu}$. Gauging this symmetry we introduce
\begin{equation}
D_{\mu}A_{\nu\rho}\equiv\partial_{\mu}A_{\nu\rho}+V_{\mu\nu\rho}.
\end{equation}
Then, the action becomes
\begin{equation}
S_{A,V}=\frac{1}{3!}\int d^{4}xD_{[\mu}A_{\nu\rho]}D^{[\mu}A^{\nu\rho]}
\end{equation}
with $\delta V_{\mu\nu\rho}=-\frac{1}{3!}\partial_{[\mu}\epsilon_{\nu\rho]}$. Again, we ensure that $V_{\mu\nu\rho}$ is pure gauge by adding
\begin{equation}
\tilde{S}_{W}=a\int d^{4}x\phi{~}^{*}W,
\end{equation}
where ${~}^{*}W\equiv\epsilon^{\mu\nu\rho\sigma}W_{\mu\nu\rho\sigma}$ and $W_{\mu\nu\rho\sigma}\equiv\partial_{[\mu}V_{\nu\rho\sigma]}$. Varying the action with respect to $\phi$ yields
\begin{equation}
\delta{\phi} :~~~~~^{*}W=0\Rightarrow W_{\mu\nu\rho\sigma}=0
\end{equation}
so that $V_{\nu\rho\sigma}=\partial_{[\nu}\lambda_{\rho\sigma]}$. We may choose a gauge $\lambda_{\rho\sigma}=0$ or otherwise redefine $\tilde{A}_{\mu\nu}=A_{\mu\nu}+3! \lambda_{\mu\nu}$ to recover $S_{A}$.

Integrating out $V_{\mu\nu\rho}$ instead we have:
\begin{equation}
\delta V_{\mu\nu\rho}:~~~~~D^{[\mu}A^{\nu\rho]}=\frac{4!}{2}a\epsilon^{\sigma\mu\nu\rho}\partial_{\rho}\phi
\end{equation}
so that
\begin{equation}
V^{\mu\nu\rho}=2a\epsilon^{\sigma\mu\nu\rho}\partial_{\sigma}\phi-\frac{1}{3!}\partial^{[\mu}A^{\nu\rho]}.
\end{equation}
Putting this expression into the action $S_{A,V}+\tilde{S}_{W}$ yields
\begin{equation}
S_{A,V}+\tilde{S}_{W}\rightarrow 3!4!a^2\int d^{4}x\partial_{\mu}\phi\partial^{\mu}\phi .
\end{equation}
So, choosing $a=\frac{\sqrt{2}}{4!}$, the action $S_{A,V}+\tilde{S}_{W}\rightarrow S_{\phi}$.

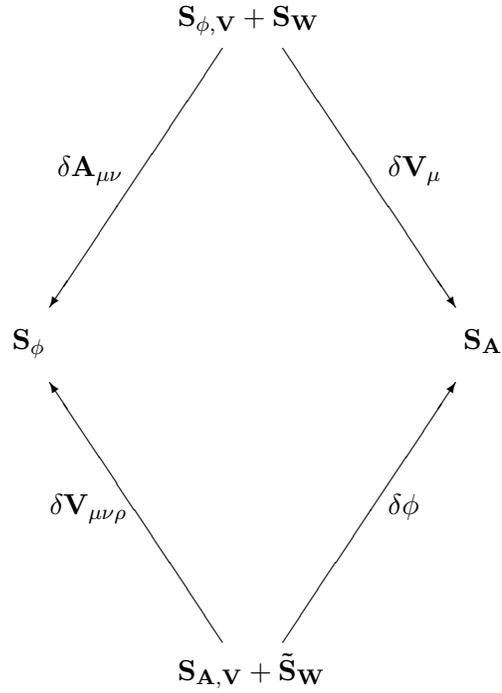
\begin{figure}[h]
\begin{center}
\setlength{\unitlength}{1mm}
\begin{picture}(100,100)
\put(40,90){$\bf{S}_{\bf{\phi} ,\bf{V}}+\bf{S_{W}}$}
\put(46,87){\vector(-2,-3){23}}\put(54,87){\vector(2,-3){23}}
\put(24,70){$\bf{\delta A_{\mu\nu}}$}\put(68,70){$\bf{\delta V_{\mu}}$}
\put(18,47){$\bf{S_{\bf{\phi}}}$}\put(78,47){$\bf{S_{A}}$}
\put(46,8){\vector(-2,3){23}}\put(54,8){\vector(2,3){23}}
\put(23,25){$\bf{\delta V_{\mu\nu\rho}}$}\put(68,25){$\bf{\delta\phi}$}
\put(40,3){$\bf{S}_{\bf{A,V}}+\bf{\tilde{S}_{W}}$}
\end{picture}
\caption{\small When the shift symmetry is gauged the two parent actions $S_{\phi ,V}+S_{W}$ and $S_{A,V}+\tilde{S}_{W}$ are (classically) equivalent to both $S_{\phi}$ and $S_{A}$. Thus $S_{\phi}$ is dual of $S_{A}$ and vice versa.}
\label{dual1-4}
\end{center}
\end{figure}

We summarize the two ways of constructing scalar-tensor duality in Figs.\ref{dual1-3} and \ref{dual1-4}:
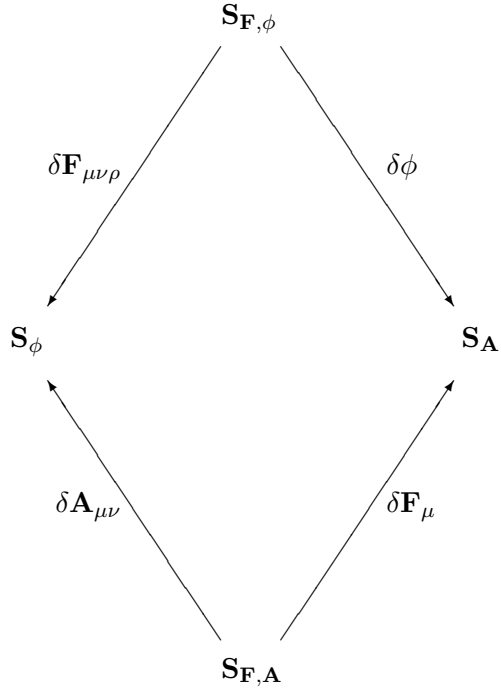
\begin{figure}[h]
\begin{center}
\setlength{\unitlength}{1mm}
\begin{picture}(100,120)
\put(46,110){$\bf{S}_{\bf{F} ,\bf{\phi}}$}
\put(46,107){\vector(-2,-3){23}}\put(54,107){\vector(2,-3){23}}
\put(23,90){$\bf{\delta F_{\mu\nu\rho}}$}\put(68,90){$\bf{\delta\phi}$}
\put(18,67){$\bf{S_{\bf{\phi}}}$}\put(78,67){$\bf{S_{A}}$}
\put(46,28){\vector(-2,3){23}}\put(54,28){\vector(2,3){23}}
\put(24,45){$\bf{\delta A_{\mu\nu}}$}\put(68,45){$\bf{\delta F_{\mu}}$}
\put(46,23){$\bf{S}_{\bf{F} ,\bf{A}}$}
\end{picture}
\caption{\small When the field equations and the Bianchi identities are interchanged the two parent actions $S_{F,\phi}$ and $S_{F,A}$ are (classically) equivalent to both $S_{\phi}$ and $S_{A}$. Thus $S_{\phi}$ is dual of $S_{A}$ and vice versa.}
\label{dual1-3}
\end{center}
\end{figure}
\pagebreak

{\bf Excercise:} Verify the important property that the abelian duality transformation $D$, is idempoten, i.e. $D^2=\pm1$, in the examples above.

\chapter{Electric-Magnetic Duality}

\section{Field eqn's $\leftrightarrow$ Bianchi's}

We now want to extend the scalar-tensor duality just described to Maxwell's theory.

The equations of motion reads
\begin{equation}
\partial_{\mu}F^{\mu\nu}=0
\end{equation}
and the Bianchi identity
\begin{equation}
\partial_{\mu}^{~~*}F^{\mu\nu}=0,
\end{equation}
in the absence of sources.

Here, interchanging field equations and the Bianchi identity is equivalent to
\begin{equation}
F^{\mu\nu}\rightarrow{~}^{*}F^{\mu\nu};~~~~~^{*}F^{\mu\nu}\rightarrow -F^{\mu\nu}
\end{equation}
Since
\begin{equation}
F^{\mu\nu}=
\begin{pmatrix} 0      & E_{x}  & E_{y}  & E_{z} \\
		-E_{x} & 0      & B_{z}  & -B_{y} \\
		-E_{y} & -B_{z} & 0      & B_{x} \\
		-E_{z} & B_{y}  & -B_{x} & 0
\end{pmatrix}
\end{equation}
and
\begin{equation}
{~}^{*}F^{\mu\nu}=
\begin{pmatrix} 0      & B_{x}  & B_{y}  & B_{z} \\
		-B_{x} & 0      & -E_{z} & E_{y} \\
		-B_{y} & E_{z}  & 0      & -E_{x} \\
		-B_{z} & -E_{y} & E_{x}  & 0
\end{pmatrix}
\end{equation}
this is tantamount to the discrete symmetry
\begin{equation}
\bf{E}\rightarrow \bf{B};~~~~~\bf{B}\rightarrow -\bf{E} \label{exch-EB}
\end{equation}
which is why this kind of duality is called electric-magnetic. 

Starting from the action
\begin{equation}
S_{A}=\frac{1}{4g^{2}}\int d^{4}xF_{\mu\nu}(A)F^{\mu\nu}(A),
\end{equation}
where $F_{\mu\nu}(A)\equiv\partial_{[\mu}A_{\nu]}$, and the Bianchi identity
\begin{equation}
\partial_{\mu}^{~~*}F^{\mu\nu}(A)=0
\end{equation}
we go to a parent action
\begin{equation}
S_{F,\Lambda}=\int d^{4}x\left(\frac{1}{4g^2}F_{\mu\nu}F^{\mu\nu}+a\Lambda_{\mu}\partial_{\nu}^{~~*}F^{\nu\mu}\right)
\end{equation}
Varying the action with respect to $\Lambda_{\mu}$ gives
\begin{equation}
\partial_{\mu}{~}^{*}F^{\mu\nu}=0\Rightarrow F_{\mu\nu}=\partial_{[\mu}A_{\nu]}
\end{equation}
so that $S_{F,\Lambda}\rightarrow S_{A}$.

On the other hand varying with respect to $F_{\mu\nu}$ gives
\begin{equation}
\delta F_{\mu\nu}:~~~~~\frac{1}{2g^2} F^{\mu\nu}=\frac{a}{2}\partial_{\rho}\Lambda_{\sigma}\epsilon^{\rho\sigma\mu\nu}\equiv\frac{a}{2}{~}^{*}G^{\mu\nu}.
\end{equation}
Plugging this back into the action, yields
\begin{equation}
S_{F,\Lambda}\rightarrow S_{\Lambda}=-\frac{g^2a^2}{4}\int d^{4}x{~}^{*}G_{\mu\nu}{~}^{*}G^{\mu\nu}
\end{equation}
Since ${~}^{*}G_{\mu\nu}{~}^{*}G^{\mu\nu}=-2G_{\mu\nu}G^{\mu\nu}$, where $G_{\mu\nu}=\partial_{[\mu}\Lambda_{\nu]}$, we obtain the dual action
\begin{equation}
S_{\Lambda}=\frac{g^2}{4}\int d^{4}xG_{\mu\nu}(\Lambda)G^{\mu\nu}(\Lambda),
\end{equation}
if\footnote{Since we are considering a free theory at the classical level the coupling constant $g$ could be scaled into the fields, and we might a priori have allowed $g$-dependence in $a$. However, the present assignment give the right charges and couplings when magnetic monopoles are considered. See \cite{kn:RAJ}.} $a=\frac{1}{\sqrt{2}}$. We see that the duality exchanged the ``coupling regimes'' $g\rightarrow g'=1/g$. Also, the gauge field $A_{\mu}$ is interchanged with $\Lambda_{\mu}$. Hence, electric-magnetic duality in 4D is a {\it vector-vector duality}.  

Now, this nice duality is destroyed when coupling to sources, unless we include magnetic ones. In fact
\begin{equation*}
\begin{alignat}{2} \bf{\nabla}\cdot\bf{E}&=\rho_{e}, &
               \qquad\bf{\nabla}\times\bf{E}&=-\frac{\partial\bf{B}}{\partial t}-{\bf{j}}_{m},\\
               \bf{\nabla}\cdot\bf{B}&=\rho_{m}, &
               \qquad\bf{\nabla}\times\bf{B}&=\frac{\partial\bf{E}}{\partial t}+{\bf{j}}_{e}
\end{alignat}
\end{equation*}
is invariant under (in complex notation)
\begin{eqnarray}
{\bf{E}}+i{\bf{B}}&{\rightarrow}& e^{i\phi}({\bf{E}}+i{\bf{B}}) \nonumber \\
\rho_{e}+i\rho_{m}&{\rightarrow}& e^{i\phi}(\rho_{e}+i\rho_{m}) \\
{\bf{j}}_{e}+i{\bf{j}}_{m}&{\rightarrow}& e^{i\phi}({\bf{j}}_{e}+i{\bf{j}}_{m}) \nonumber \label{complex-transf}
\end{eqnarray}
So, it looks as if we have an even larger group than previously contemplated in (\ref{exch-EB}). However, if we have particles with electric and magnetic charge the transformations (\ref{complex-transf}) must be accompanied by
\begin{equation}
q_{e}+iq_{m}\rightarrow e^{i\phi}(q_{e}+iq_{m}).
\end{equation}
But, the Dirac quatization rule says\footnote{Here $i,j$ denumbers the charges of the different particles.}
\begin{equation}
q_{e}^{i}q_{m}^{j}=2\pi\hbar\eta_{ij}
\end{equation}
where $\eta_{ij}$ is a matrix of integers. So we have invariance if $\phi=\pm\frac{\pi}{2}$.

\section{p-form duality}

Let us now collect and generalize what we have learned so far.

Suppose we study a {\it p-form theory in D dimensions}, i.e.
\begin{equation}
\begin{split}
A&=\frac{1}{p!}A_{\mu_{1}\mu_{2}\dots\mu_{p}}dx^{\mu_{1}}\wedge dx^{\mu_{2}}\dots\wedge dx^{\mu_{p}}  \\
F{\equiv}dA&=\frac{1}{(p+1)!}\partial_{[\mu_{1}}A_{\mu_{2}\mu_{3}\dots\mu_{p+1}]}dx^{\mu_{1}}\wedge\dots\wedge dx^{\mu_{p+1}},
\end{split}
\end{equation}
where $F$ is the field strength; $F_{\mu_{1}\mu_{2}\dots\mu_{p+1}}=\partial_{[\mu_{1}}A_{\mu_{2}\dots\mu_{p+1}]}$ and $p\leqslant D-2$.

The action is
\begin{eqnarray}
S_{A}&=&\frac{1}{2(p+1)}\int d^{D}xF_{\mu_{1}\mu_{2}\dots\mu_{p+1}}(A)F^{\mu_{1}\mu_{2}\dots\mu_{p+1}}(A) \nonumber \\
&=&\frac{1}{2(p+1)}\int d^{D}xF^2(A)
\end{eqnarray}
So the field equations and the Bianchi identities are, respectively
\begin{eqnarray}
\partial_{\mu}F^{\mu\mu_{1}\dots\mu_{p}}(A)&=&0 \\
\partial_{\mu}{~}^{*}F^{\mu\mu_{1}\dots\mu_{D-(p+2)}}(A)&=&0
\end{eqnarray}

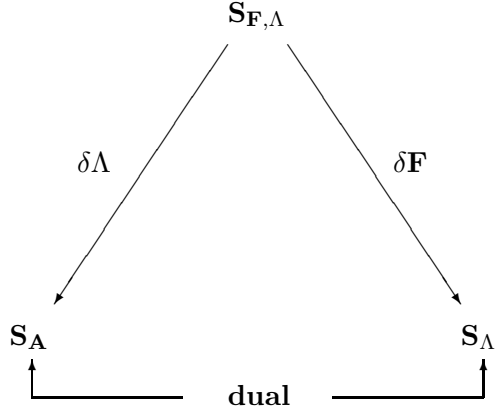
\begin{figure}
\begin{center}
\setlength{\unitlength}{1mm}
\begin{picture}(100,70)
\put(46,60){$\bf{S}_{\bf{F},\bf{\Lambda}}$}
\put(46,57){\vector(-2,-3){23}}\put(54,57){\vector(2,-3){23}}
\put(26,40){$\bf{\delta \Lambda}$}\put(68,40){$\bf{\delta F}$}
\put(17,17){$\bf{S}_{\bf{A}}$}\put(77,17){$\bf{S}_{\bf{\Lambda}}$}
\put(20,10){\vector(0,1){5}}\put(80,10){\vector(0,1){5}}
\put(20,10){\line(1,0){20}}\put(60,10){\line(1,0){20}}
\put(46,9){\bf{dual}}
\end{picture}
\caption{\small The parent action $S_{F,\Lambda}$ is (classically) equivalent to both $S_{A}$ and $S_{\Lambda}$ showing that $S_{A}$ is dual to $S_{\Lambda}$ and vice versa.}
\label{dual2-1}
\end{center}
\end{figure}

A parent action is e.g.
\begin{equation}
S_{F,\Lambda}=\frac{1}{2(p+1)}\int d^{D}x\left(F^2+a\Lambda\partial^{*}F\right)
\end{equation}
where $\Lambda$ is a $D-(p+2)$-form. Varying this action with respect to $\Lambda$ get back $S_{A}$. If we instead vary the parent action with respect to $F$ and put the expression for $F$ back into $S_{F,\Lambda}$, the parent action transforms into the dual action
\begin{equation}
S_{\Lambda}=\frac{1}{2(p+1)}\int d^{D}xF^{2}(\Lambda)
\end{equation}
where $F(\Lambda)$ is a $D-(p+1)$-form ($F(\Lambda)\sim~^{*}F(A)$). Hence, there is a dualization between the $p$-form $A$ and the $D-(p+2)$-form $\Lambda$. The diagram is now (Fig.\ref{dual2-1})

In the picture where we gauge the shift-symmetry the situation looks as follows. The action
\begin{eqnarray}
S_{A}&=&\frac{1}{2(p+1)}\int d^{4}xF^{2}(A) \nonumber \\
&=&\frac{1}{2(p+1)}\int d^{4}x\partial A\partial A
\end{eqnarray}
is invariant under the global symmetry $A\rightarrow A+\epsilon$. When we gauge this symmetry the action changes into
\begin{equation}
S_{A,V}=\frac{1}{2(p+1)}\int d^{4}xDADA
\end{equation}
where $DA=\partial A+V$ and where $V$ is a $p+1$-form.
The parent action is
\begin{equation}
S_{A,V,\Lambda}=\frac{1}{2(p+1)}\int d^{D}x\left(DADA+\Lambda^{~*}W\right)
\end{equation}
where $\Lambda$ and ${~}^{*}W$ are $D-(p+2)$-forms. This is summarized in Fig.\ref{dual2-2} below. 
\begin{figure}
\begin{center}
\setlength{\unitlength}{1mm}
\begin{picture}(100,70)
\put(44,60){$\bf{S}_{\bf{A},\bf{V},\bf{\Lambda}}$}
\put(46,57){\vector(-2,-3){23}}\put(54,57){\vector(2,-3){23}}
\put(26,40){$\bf{\delta \Lambda}$}\put(68,40){$\bf{\delta V}$}
\put(17,17){$\bf{S}_{\bf{A}}$}\put(77,17){$\bf{S}_{\bf{\Lambda}}$}
\put(20,10){\vector(0,1){5}}\put(80,10){\vector(0,1){5}}
\put(20,10){\line(1,0){20}}\put(60,10){\line(1,0){20}}
\put(46,9){\bf{dual}}
\end{picture}
\caption{\small The parent action $S_{A,V,\Lambda}$ is (classically) equivalent to both $S_{A}$ and $S_{\Lambda}$, showing that $S_{A}$ is dual to $S_{\Lambda}$ and versa vice.}
\label{dual2-2}
\end{center}
\end{figure}
Using this language we collect our previously discussed $D=4$ examples in the table \ref{table2-1}. 
\begin{table}
\begin{center}
\begin{tabular}{|l|l|l|l|l|}                      \hline
$\bf{p}$ & \multirow{1}{6mm}{$\bf{A_{p}}$} & \multirow{1}{13mm}{$\bf{F_{p+1}}$} & \multirow{1}{16mm}{$\bf{\Lambda_{D-(2+p)}}$} & \multirow{1}{19mm}{$\bf{F(\Lambda)_{D-p-1}}$} \\ \hline\hline
$0$ & \multirow{1}{6mm}{$A$} & \multirow{1}{13mm}{$F_{\mu}(A)$} & \multirow{1}{16mm}{$\Lambda_{\mu\nu}$} & \multirow{1}{19mm}{$F_{\mu\nu\rho}(\Lambda)$} \\ \hline
1 & \multirow{1}{6mm}{$A_{\mu}$} & \multirow{1}{13mm}{$F_{\mu\nu}(A)$} & \multirow{1}{16mm}{$\Lambda_{\mu}$} & \multirow{1}{19mm}{$F_{\mu\nu}(\Lambda)$} \\ \hline
2 & \multirow{1}{6mm}{$A_{\mu\nu}$} & \multirow{1}{13mm}{$F_{\mu\nu\rho}(A)$} & \multirow{1}{16mm}{$\Lambda$} & \multirow{1}{19mm}{$F_{\mu}(\Lambda)$} \\ \hline
\end{tabular}
\caption{\small The various possibilities of p-form dualities in $D=4$.}
\label{table2-1}
\end{center}
\end{table} 
 Here we have renamed the scalar field $\phi$ and called it $A$ viewing it as a zero form. In $D=3$ the corresponding table is seen in table \ref{table2-2}.
\begin{table}                          
\begin{center}
\begin{tabular}{|l|l|l|l|l|}                      \hline
$\bf{p}$ & \multirow{1}{6mm}{$\bf{A_{p}}$} & \multirow{1}{13mm}{$\bf{F_{p+1}}$} & \multirow{1}{16mm}{$\bf{\Lambda_{D-(2+p)}}$} & \multirow{1}{19mm}{$\bf{F(\Lambda)_{D-p-1}}$} \\ \hline\hline
$0$ & \multirow{1}{6mm}{$A$} & \multirow{1}{13mm}{$F_{\mu}(A)$} & \multirow{1}{16mm}{$\Lambda_{\mu}$} & \multirow{1}{19mm}{$F_{\mu\nu}(\Lambda)$} \\ \hline
1 & \multirow{1}{6mm}{$A_{\mu}$} & \multirow{1}{13mm}{$F_{\mu\nu}(A)$} & \multirow{1}{16mm}{$\Lambda$} & \multirow{1}{19mm}{$F_{\mu}(\Lambda)$} \\ \hline
\end{tabular}
\caption{\small The various possibilities of p-form dualities in $D=3$.}
\label{table2-2}
\end{center}
\end{table}
\pagebreak
\\ 

{\it \bf Excercise:} Construct the various parent actions corresponding to table \ref{table2-2}.

\section{3D vector-vector duality}

\subsection{Dualization between a self dual vector field and a self dual topologically massive vector gauge field}

To illustrate that there may be more complicated dualities than the p-form dualities just described, let us look at a 3D example \cite{kn:Des}.

The action (which is linear in derivatives)
\begin{equation}
S_{B}=\frac{1}{2}\int d^{3}x\left(m^2B_{\mu}B^{\mu}-\frac{1}{2}m\epsilon^{\mu\nu\rho}B_{\mu}\partial_{\nu}B_{\rho}\right) \label{act-B}
\end{equation}
describes a {\it massive, self dual vector field}. The field equation reads
\begin{equation}
\delta B_{\mu}:~~~~~2m^{2}B^{\mu}-m\epsilon^{\mu\nu\rho}\partial_{\nu}B_{\rho}=0.
\end{equation}
Defining $F_{\mu\nu}(B)\equiv\partial_{[\nu}B_{\rho]}$ and $~^{*}F^{\mu}\equiv\frac{1}{2}\epsilon^{\mu\nu\rho}F_{\nu\rho}$ we get {\it the self duality condition}\footnote{Here ``dual'' refers to Hodge duality.}: 
\begin{equation}
B^{\mu}=\frac{1}{2m}{~}^{*}F^{\mu}(B)
\end{equation}

The action (\ref{act-B}) is in fact dual to {\it another} action for a {\it self dual topologically massive vector gauge field} $A_{\mu}$ with action
\begin{equation}
S_{A}=\int d^{3}x\left(-\frac{1}{4}F_{\mu\nu}(A)F^{\mu\nu}(A)+\frac{1}{2}m\epsilon^{\mu\nu\rho}A_{\mu}F_{\nu\rho}(A)\right)
\end{equation}
which is quadratic in derivatives; $F_{\mu\nu}(A)\equiv\partial_{[\mu}A_{\nu]}$. The field equations are
\begin{equation}
\delta A_{\mu}:~~~~~\partial_{\mu}F^{\mu\nu}(A)+m\epsilon^{\mu\nu\rho}F_{\nu\rho}=0
\end{equation}
giving {\it the self duality condition}: 
\begin{equation}
\partial_{\mu}F^{\mu\nu}=-2m{~}^{*}F^{\nu}
\end{equation}
The duality can, as we are accustomed to by now, be seen in many different ways. One parent action is
\begin{equation}
S_{B,A}=\frac{1}{2}\int d^{3}x\left\{m^{2}B_{\mu}B^{\mu}+m\epsilon^{\mu\nu\rho}\left(B_{\mu}F_{\nu\rho}(A)+A_{\mu}F_{\nu\rho}(A)\right)\right\} \label{3D-parent}
\end{equation}

The equivalence to $S_{A}$ is immediately, the field equation,
\begin{equation}
\delta B_{\mu}:~~~~~B^{\mu}=-\frac{1}{2m}\epsilon^{\mu\nu\rho}F_{\nu\rho}(A)
\end{equation}
returns $S_{A}$ when plugged back into $S_{B,A}$. On the other hand
\begin{equation}
\delta A_{\mu}:~~~~~\epsilon^{\mu\nu\rho}\partial_{\nu}B_{\rho}+\epsilon^{\mu\nu\rho}F_{\nu\rho}(A)=0 \label{eq-A}
\end{equation}
returns $S_{B}$ if we partially integrate in the {\it Chern-Simons} term and use the field equations (\ref{eq-A}) twice:
\begin{eqnarray}
S_{B,A}&\rightarrow &\int d^3x\{m^2B_{\mu}B^{\mu}-m\epsilon^{\mu\nu\rho}(B_{\mu}\partial_{\nu}B_{\rho}+A_{\mu}\partial_{\nu}B_{\rho})\} \nonumber \\
&=&\int d^3x\{m^2B_{\mu}B^{\mu}-m\epsilon^{\mu\nu\rho}(B_{\mu}\partial_{\nu}B_{\rho}-B_{\rho}\partial_{\nu}A_{\mu})\} \nonumber \\
&=&\int d^3x\{m^2B_{\mu}B^{\mu}-\frac{1}{2}m\epsilon^{\mu\nu\rho}B_{\mu}\partial_{\nu}B_{\rho}\}
\end{eqnarray}

Another parent action is
\begin{equation}
S_{A,B}=\int d^{3}x\left\{-\frac{1}{4}F_{\mu\nu}(A)F^{\mu\nu}(A)+\frac{1}{2}m\epsilon^{\mu\nu\rho}\left(F_{\mu\nu}(A)B_{\rho}-\frac{1}{4}F_{\mu\nu}(B)B_{\rho}\right)\right\}
\end{equation}
Varying $S_{A,B}$ with respect to $B_{\mu}$ gives back $S_{A}$ and $S_{B}$ is obtained when $S_{A,B}$ is varied with respect to $A_{\mu}$.

\subsection{3D dualization from 2D point of view}

Let us dimensionally reduce the above models. Thus, consider again the 3D parent action (\ref{3D-parent})
\begin{equation*}
S_{B,A}=\frac{1}{2}\int d^{3}x\left\{m^2B_{\mu}B^{\mu}+m\epsilon^{\mu\nu\rho}(B_{\mu}F_{\nu\rho}(A)+A_{\mu}F_{\nu\rho}(A))\right\}
\end{equation*}
Then make a 2+1 split of the vector fields
\begin{equation}
B_{\mu}\rightarrow(B_{\mu},\phi);~~~~~A_{\mu}\rightarrow(A_{\mu},\lambda)
\end{equation}
Assuming the fields have no dependence on the third coordinate the 2D parent action may be written (up to boundary terms)
\begin{equation}
S_{B,\phi,A,\lambda}=\frac{1}{2}\int d^{2}x\left\{m^2(B_{\mu}B^{\mu}+\phi^2)+2m\epsilon^{\mu\nu}(B_{\mu}\partial_{\nu}\lambda+\phi\partial_{\mu}A_{\nu}+2\lambda\partial_{\mu}A_{\nu})\right\} \label{3D-parent2}
\end{equation}
Varying the 2D parent action with respect to $B_{\mu}$ and $\phi$ yields
\begin{equation}
\begin{split}
\delta B_{\mu}&:~~~~~2m^2B^{\mu}+2m\epsilon^{\mu\nu}\partial_{\nu}\lambda=0~~~\Rightarrow ~~~B^{\mu}=-\frac{1}{m}\epsilon^{\mu\nu}\partial_{\nu}\lambda \\
\\
\delta\phi &:~~~~~2m^2\phi+2m\epsilon^{\mu\nu}\partial_{\mu}A_{\nu}=0~~~\Rightarrow ~~~\phi=-\frac{1}{m}\epsilon^{\mu\nu}\partial_{\mu}A_{\nu} \\
\end{split}
\end{equation}
Putting these expressions into $S_{B,\phi,A,\lambda}$ gives the action
\begin{equation}
S_{A, \lambda}=\int d^{2}x\left\{-\frac{1}{4}F_{\mu\nu}F^{\mu\nu}+\frac{1}{2}\partial_{\mu}\lambda\partial^{\mu}\lambda +m\epsilon^{\mu\nu}\lambda F_{\mu\nu}\right\} 
\end{equation}
where $F_{\mu\nu}\equiv\partial_{[\mu}A_{\nu]}$. $S_{A,\lambda}$ is dual to another action $S_{B,\phi}$ which is found by varying the parent action (\ref{3D-parent2}) with respect to $A_{\mu}$ and $\lambda$. So it is a simultaneous dualization of a vector and a scalar which should be interpretable as two scalar-scalar dualities in 2D. 

Another parent action is
\begin{eqnarray}
S_{A,\lambda ,B,\phi}&=&\int d^{2}x\left\{-\frac{1}{4}F_{\mu\nu}(A)F^{\mu\nu}(A)-\frac{1}{2}\partial_{\mu}\lambda\partial^{\mu}\lambda \right. \nonumber \\ \nonumber \\
& & \qquad\qquad \left. -m\epsilon^{\mu\nu}\left(-B_{\mu}\partial_{\nu}\lambda-\phi\partial_{\mu}A_{\nu}+\frac{1}{2}\phi\partial_{\mu}B_{\nu}\right)\right\}
\end{eqnarray}

Again, dualization proceeds as in the previous case. Note that we have more possibilities in $2D$ than in $3D$, however. When we relax the requirement that $(B_{\mu},\phi)$ or $(A_{\mu},\lambda)$ (the $3D$ vectors) should be integrated out we may e.g. integrate out $B_{\mu}$ and $\lambda$, say.

In \cite{kn:LKRN}, a non-Abelian versions of the $3D$ duality discussed here was given. It might be interesting to look at the $2D$ non-Abelian dualities that arise from those models. 

Now, what about non-abelian theories in general? Due to the self interaction we do not have a shift symmetry to gauge and due to the non-abelian nature, the Bianchi identity is not directly integrable (Rememeber $D_{\mu}$ contains the connection). It is therefore a bit interesting that the above 3D duality can be generalized to non-abelian form. It is one of very few such examples.  

For $\sigma$-model duality there exists a systematic non-abelian dualization however. We shall return to the question in that context.
\\

{\it \bf Exercise:} Find the remaining dual actions in $2D$.

\chapter{$\sigma$-Model Duality}

\section{String interlude} \label{string-int}

String theories are based on 2D non-linear $\sigma$-models with bosonic (and fermionic) degrees of freedom. The bosonic part of the critical superstring action may be written (with no coupling to background fields)
\begin{equation}
S=-\frac{1}{4\pi\alpha '}\int d^{2}\xi\sqrt{-g}g^{ab}\gamma_{ab}
\end{equation}
where $\alpha '$ is the inverse string tension, $g_{ab}$ is an auxiliary metric on the world sheet and the induced metric $\gamma_{ab}$ is
\begin{equation}
\gamma_{ab}=\partial_{a}X^{\mu}\partial_{b}X^{\nu}\eta_{\mu\nu},
\end{equation}
$\eta_{\mu\nu}$ is the trivial background\footnote{We use $a,b\ldots =0,1$ as world sheet indices and $\mu ,\nu\ldots =0,1,\ldots ,9$ as space-time indices.}.
 
The coordinates $X^{\mu}$ on $M^{10}$ (target space) are given by a mapping
\begin{equation}
X^{\mu}:~~~M^{2}\longrightarrow M^{10}
\end{equation}

The string action including coupling to background fields is given in covariant gauge by
\begin{equation}
S=-\frac{1}{4\pi\alpha '}\int d^{2}\xi\left\{\partial_{a}X^{\mu}\partial^{a}X^{\nu}G_{\mu\nu}(X)+\epsilon^{ab}\partial_{a}X^{\mu}\partial_{b}X^{\nu}B_{\mu\nu}(X)-\alpha 'R^{(2)}\phi(X)\right\}
\end{equation}
where $G_{\mu\nu}$ is the non-trivial background metric, $B_{\mu\nu}$ is an antisymmetric tensor field, $\phi$ a scalar field and $R^{(2)}$ is the world-sheet Ricci scalar. This action has a duality called $T$-duality (target space duality).

To lowest order in the string parameter $\alpha '$ the vanishing of the $\beta$-functions, i.e. the requirement of scale invariance of the quantum theory, results in field equations for the background geometry fields $G$, $B$ and $\phi$ that may be summarized in the effective action
\begin{equation}
S_{\mbox{eff}}=-\frac{1}{2\kappa^2}\int d^{10}X\sqrt{-G}e^{-2\phi}\left\{R^{(10)}+\frac{1}{6}F_{\mu\nu\rho}F^{\mu\nu\rho}-4D_{\mu}\phi D^{\mu}\phi\right\}. \label{low-energy}
\end{equation}
Here $\kappa^2$ is the gravitational coupling, $R^{(10)}$ is the Ricci curvature scalar of the ten-dimensional target space (space-time), $F_{\mu\nu\rho}$ is the field strength of $B_{\mu\nu}$ ($F_{\mu\nu\rho}=\partial_{[\mu}B_{\nu\rho]}$) and $D_{\mu}$ is the covariant derivatives on the target space. The action (\ref{low-energy}) has a $S$-duality\footnote{S duality is the kind of duality we treated in the previous sections.} symmetry. When supersymmetry is taken into account more fields are needed.

Amongst the string dualities one also find $U$-duality which is a combination of $T$- and $S$-duality, including dimensional reduction \cite{kn:HullT}. We will not discuss $U$-duality in these notes.

We now turn to the second type of duality that is relevant for String Theory. $T$-duality is a transformation that acts on 2D sigma-models. Before we rush into $T$-duality, we first present some preliminaries on $D$-dimensional $\sigma$-models and construct the dual action when the target space has an isometry. 

\section{$\sigma$-models, target space isometries and the\\ Legendre transform} \label{isometries}

A $\sigma$-model is a map from a $d$-dimensional space $M$ into a $D$-dimensional target space $T$
\begin{equation}
\phi^{\mu}:~~M\longrightarrow T
\end{equation}
with action
\begin{equation}
S_{\phi}=\int d^{d}\xi\partial_{a}\phi^{\mu}\partial^{a}\phi^{\nu}G_{\mu\nu}(\phi) \label{sigma-action}
\end{equation}
where $\partial^{a}\equiv\eta^{ab}\partial_{b}$, $\eta_{ab}$ is the Minkowski metric and $\partial_{a}\equiv\frac{\partial}{\partial\xi^{a}}$.

With $G_{\mu\nu}$ nontrivial, it is a non-linear $\sigma$-model. $G_{\mu\nu}$ has the interpretation of a metric on $T$. The equation of motion that follows from (\ref{sigma-action}) reads
\begin{equation}
\delta\phi^{\mu}:~~~~~\partial_{a}[\partial^{a}\phi^{\nu}G_{\mu\nu}(\phi)]-\frac{1}{2}G_{\rho\sigma ,\mu}\partial_{a}\phi^{\rho}\partial^{a}\phi^{\sigma}=0 \label{field-eq}
\end{equation}
As an aside, we rewrite this as
\begin{eqnarray}
0&=&\partial^{2}\phi^{\nu}G_{\mu\nu}+\partial_{a}\phi^{\rho}\partial^{a}\phi^{\nu}G_{\mu\nu ,\rho}-\frac{1}{2}G_{\rho\sigma ,\mu}\partial_{a}\phi^{\rho}\partial^{a}\phi^{\sigma} \nonumber \\
&=&\partial^{2}\phi^{\nu}G_{\mu\nu}+\frac{1}{2}\partial_{a}\phi^{\rho}\partial^{a}\phi^{\sigma}(2G_{\mu\sigma ,\rho}-G_{\sigma\rho ,\mu}) \nonumber \\
&\Rightarrow &\partial^{2}\phi^{\mu}+\frac{1}{2}G^{\mu\nu}(G_{\nu\sigma ,\rho}+G_{\nu\rho ,\sigma}-G_{\sigma\rho ,\nu})\partial_{a}\phi^{\rho}\partial^{a}\phi^{\sigma}=0, \nonumber
\end{eqnarray}
so that
\begin{equation}
\partial^{2}\phi^{\mu}+\Gamma^{\mu}_{~~\sigma\rho}\partial_{a}\phi^{\rho}\partial^{a}\phi^{\sigma}=0
\end{equation}
where the Levi-Civita connection is
\begin{equation}
\Gamma^{\mu}_{~~\sigma\rho}\equiv\frac{1}{2}G^{\mu\nu}(G_{\nu\sigma ,\rho}+G_{\nu\rho ,\sigma}-G_{\sigma\rho ,\nu})
\end{equation}
The pullback $D_{a}$ to $M$ of the covariant derivative $D_{\mu}$ on $T$ is
\begin{equation}
D_{a}\equiv\partial_{a}\phi^{\mu}D_{\mu}
\end{equation}
with
\begin{equation}
D_{\mu}=\partial_{\mu}+\Gamma_{\mu}
\end{equation}
Now $\phi^{\mu}$ is a coordinate on $T$, so it is not a vector, but $\partial_{a}\phi^{\mu}$ is, since it involves the difference $\Delta\phi^{\mu}$. So we may write
\begin{eqnarray}
D^{a}\partial_{a}\phi^{\mu}&=&\partial^{a}\phi^{\rho}(\partial_{\rho}\partial_{a}\phi^{\mu}+\Gamma_{\rho\sigma}^{~~~\mu}\partial_{a}\phi^{\sigma}) \nonumber \\
&=&\partial^{2}\phi^{\mu}+\Gamma_{\rho\sigma}^{~~~\mu}\partial^{a}\phi^{\rho}\partial_{a}\phi^{\sigma}.
\end{eqnarray}
Hence our field equation (\ref{field-eq}) may be written
\begin{equation}
D^{a}\partial_{a}\phi^{\mu}=0
\end{equation}
Thus the name ``harmonic map''.

Now back to the $\sigma$-model action (\ref{sigma-action})
\begin{equation*}
S_{\phi}=\int d^{d}\xi G_{\mu\nu}(\phi)\partial_{a}\phi^{\mu}\partial^{a}\phi^{\nu}
\end{equation*}
Suppose now that the target space metric $G_{\mu\nu}$ has an isometry, given by a vector field $\epsilon^{\mu}(\phi)$. Then the Lie derivative of the metric along the vector field $\epsilon^{\mu}(\phi)$ vanishes, i.e.,
\begin{equation}
\pounds_{\bf{\epsilon}}G_{\mu\nu}=0~~~\Leftrightarrow ~~~\delta_{\bf{\epsilon}}G_{\mu\nu}=0,~~\delta_{\bf{\epsilon}}\phi^{\mu}=\epsilon^{\mu}.
\end{equation}

We first show that isometry is a symmetry of the $\sigma$-model action:
\begin{eqnarray}
\delta_{\bf{\epsilon}}S&=&\int d^{d}\xi\left\{G_{\mu\nu,\rho}\epsilon^{\rho}\partial_{a}\phi^{\mu}\partial^{a}\phi^{\nu}+2G_{\mu\nu}\partial_{a}\epsilon^{\mu}\partial^{a}\phi^{\nu}\right\} \nonumber \\
&=&\int d^{d}\xi\{G_{\mu\nu ,\lambda}\epsilon^{\lambda}+2G_{\lambda\nu}\epsilon^{\lambda}_{~,\mu}\}\partial_{a}\phi^{\mu}\partial^{a}\phi^{\nu}. \nonumber
\end{eqnarray}
Using that
\begin{equation}
\frac{1}{2}G_{\mu(\nu}\epsilon^{\mu}_{~,\rho)}+\frac{1}{2}G_{\nu\rho ,\mu}\epsilon^{\mu}=\frac{1}{2}G_{\mu(\nu}\nabla_{\rho)}\epsilon^{\mu},
\end{equation}
we may write
\begin{eqnarray}
\delta_{\bf{\epsilon}}S&=&\int d^{d}\xi G_{\mu(\nu}\nabla_{\rho)}\epsilon^{\mu}\partial_{a}\phi^{\nu}\partial^{a}\phi^{\rho} \nonumber \\
&=&\int d^{d}\xi\nabla_{(\mu}\epsilon_{\nu)}\partial_{a}\phi^{\mu}\partial^{a}\phi^{\nu} \nonumber \\
&=&0 
\end{eqnarray}
as long as $\epsilon^{\mu}$ is a {\it Killing field}, i.e. it generates an isometry.

Choosing adapted coordinates, i.e., coordinates such that $\bf\epsilon=\frac{\partial}{\partial\phi^{0}}$ say, we find the following parent action \cite{kn:HKLR}
\begin{equation}
S_{\phi ,V,\Lambda}=\int d^{d}\xi\left(G_{00}V^{a}V_{a}+2G_{0i}V^{a}\partial_{a}\phi^{i}+G_{ij}\partial^{a}\phi^{i}\partial_{a}\phi^{j}+\Lambda^{ab}\partial_{a}V_{b}\right).
\end{equation}
Varying the $\sigma$-model action with respect to $\Lambda^{ab}$ yields
\begin{equation}
\delta\Lambda^{ab} :~~~~~\partial_{[a}V_{b]}=0
\end{equation}
Hence $V_{a}=\partial_{a}\phi^{0}$ such that $S_{\phi ,V,\Lambda}\rightarrow S_{\phi}$. Varying the action instead with respect to $V_{a}$ yields
\begin{eqnarray}
\delta V_{a}:&~~~~~&2G_{00}V^{a}+2G_{0i}\partial^{a}\phi^{i}-\partial_{b}\Lambda^{ba}=0 \nonumber \\ \nonumber \\
&\Longrightarrow &V^{a}=(G_{00})^{-1}\left[\frac{1}{2}\partial_{b}\Lambda^{ba}-G_{0i}\partial^{a}\phi^{i}\right] 
\end{eqnarray}

Putting the expression for $V_{a}$ back into the action yields
\begin{eqnarray}
S_{\phi ,V,\Lambda}\rightarrow \tilde{S}
&=&\int d^{d}x\left\{\frac{1}{G_{00}}\left[-\frac{1}{4}\partial_{b}\Lambda^{ba}\partial_{c}\Lambda^{c}_{~a}+G_{0i}\partial_{b}\Lambda^{ba}\partial_{a}\phi^{i}\right]\right. \nonumber \\ \nonumber \\
& &\qquad\qquad+\left. \left[G_{ij}-\frac{G_{i0}G_{0j}}{G_{00}}\right]\partial_{a}\phi^{i}\partial^{a}\phi^{j}\right\}. 
\end{eqnarray}

Now, $\partial_{b}\Lambda^{ba}$ may be written as a field strength of a $d-2$ form by taking the Hodge dual
\begin{equation*}
A_{a_{1}\ldots a_{d-2}}=\frac{1}{2!}\epsilon_{a_{1}\ldots a_{d-2}bc}\Lambda^{bc}~~~\Rightarrow ~~~\Lambda^{bc}=\frac{1}{(d-2)!}\epsilon^{bca_{1}\ldots a_{d-2}}A_{a_{1}\ldots a_{d-2}} 
\end{equation*}
so that
\begin{eqnarray}
\partial_{b}\Lambda^{bc}&=&\frac{1}{(d-2)!}\epsilon^{bca_{1}\ldots a_{d-2}}\partial_{b}A_{a_{1}\ldots a_{d-2}} \nonumber \\
&=&-\frac{1}{(d-2)!}\frac{1}{(d-1)!}\epsilon^{bca_{1}\ldots a_{d-2}}F_{ba_{1}a_{2}\ldots a_{d-2}}
\end{eqnarray}
where $F_{ba_{1}a_{2}\ldots a_{d-2}}\equiv\partial_{[b}A_{a_{1}a_{2}\ldots a_{d-2}]}$. Using that
\begin{equation}
~^{*}F^{c}\equiv -\frac{1}{(d-1)!}\epsilon^{bca_{1}\ldots a_{d-2}}F_{ba_{1}a_{2}\ldots a_{d-2}} \nonumber
\end{equation}
we may write
\begin{equation}
\partial_{b}\Lambda^{bc}=\frac{1}{(d-2)!}~^{*}F^{c}
\end{equation}
Furthermore we have $~^{*}F_{c}~^{*}F^{c}=(-)^{d-1}(d-1)!F_{ba_{1}a_{2}\ldots a_{d-2}}F^{ba_{1}a_{2}\ldots a_{d-2}}$. 

Again, we see the characteristic feature of duality: the field equations for the original action
\begin{equation}
\delta\phi^{0} :~~~~~\partial_{a}(G_{00}V^{a}(\phi^{0})+G_{0i}\partial^{a}\phi^{i})=0;~~~~~V^{a}(\phi^{0})\equiv\partial^{a}\phi^{0} \label{A}
\end{equation}
and the field equations from the dual action
\begin{equation}
\delta\Lambda^{ba} :~~~~~\partial_{[a}V_{b]}=0;~~~~~V_{b}(\Lambda)=G_{00}^{-1}\left(\frac{1}{2}\partial_{a}\Lambda^{a}_{~b}-G_{0i}\partial_{b}\phi^{i}\right) \label{B}
\end{equation}
are related by (\ref{B}) being the Bianchi identities for (\ref{A}) and (\ref{A}) being the Bianchi identities for (\ref{B}): {\it duality interchanges field equations and Bianchi identities}. 

The duality construction described is actually a {\it Legendre transform}\cite{kn:LR}:
\begin{eqnarray}
\partial_{a}\phi^{0}\rightarrow V_{a};~&~&~~~{\cal{L}}(V_{a})+\Lambda^{ab}\partial_{a}V_{b}=\tilde{\cal{L}}(F(A)) \nonumber \\ \nonumber \\
& &\frac{\delta\cal{L}}{\delta V_{a}}=\partial_{b}\Lambda^{ba}\sim~^{*}F^{a} \nonumber \Rightarrow V(F(A)) \\
& &\frac{\delta\tilde{\cal{L}}}{\delta F}=-~^{*}V \nonumber
\end{eqnarray}

The above construction generalizes immediately to $N$ commuting (abelian) isometries.

\section{T-duality}

Let us now descend to $d=2$. This is the dimension relevant for strings. As discussed in Section \ref{string-int}, strings moving in a nontrivial background are described by the action (in conformal gauge)
\begin{equation}
S=\int d^{2}\xi\left\{\partial^{a}X^{\mu}\partial_{a}X^{\nu}G_{\mu\nu}(X)+\epsilon^{ab}\partial_{a}X^{\mu}\partial_{b}X^{\nu}B_{\mu\nu}(X)\right\}
\end{equation}
where $\phi^{\mu}\rightarrow X^{\mu}$ to indicate the interpretation of the target space as space-time. We encounter the new possibility of having a parity breaking term involving an antisymmetric tensor field $B_{\mu\nu}$, called the Kalb-Ramond field. Physically $G_{\mu\nu}$ is the metric and $B_{\mu\nu}$ is the potential for the torsion in space-time. There is also in general a dilaton field $\phi$ which enters the action as $\int d^{2}\xi \sqrt{-g}\phi R^{(2)}$ (in a general gauge).

Assuming that there is a generalized isometry, i.e., a transformation that leaves $G$ and $B$ invariant, i.e.
\begin{equation}
\pounds_{\bf{\epsilon}}G_{\mu\nu}=\pounds_{\bf{\epsilon}}B_{\mu\nu}=0,
\end{equation}
the above $\sigma$-model action can be dualized to another one ($\sim$ denotes dual quantities)
\begin{equation}
\tilde{S}=\int d^{2}\xi\left\{\partial_{a}\tilde{X}^{\mu}\partial^{a}\tilde{X}^{\nu}\tilde{G}_{\mu\nu}(\tilde{X})+\epsilon^{ab}\partial_{a}\tilde{X}^{\mu}\partial_{b}\tilde{X}^{\nu}\tilde{B}_{\mu\nu}(\tilde{X})\right\}
\end{equation}
with the dual metric, dilaton and Kalb-Ramond field satisfying the Buscher rules \cite{kn:TB}
\begin{equation}
\begin{split}
\tilde{G}_{00}&=G^{-1}_{00} \\
\tilde{G}_{0i}&=G^{-1}_{00}B_{0i} \\
\tilde{G}_{ij}&={G}_{ij}-G^{-1}_{00}(G_{i0}G_{0j}+B_{i0}B_{0j}) \\
\tilde{B}_{ij}&=B_{ij}+G^{-1}_{00}(G_{i0}B_{0j}+B_{i0}G_{0j}) \\
\tilde{B}_{0i}&=G^{-1}_{00}G_{0i} \\
\tilde{\phi}&=\phi-\frac{1}{2}\mbox{ln}G_{00} \\
\end{split}
\end{equation}
The transformation of the dilaton field $\phi$ is a one-loop effect. It results, e.g., from a transformation of the measure if the dualization is performed in the path integral.

\section{The bosonic $O(3)$ model}

In this section we apply the dualization rules to a specific example, the $O(3)$ model. We dualize the bosonic model as well as its $N=2$ susy extension. In the latter case we use the superspace dualization prescription \cite{kn:HKLR,kn:LR,kn:GHR,kn:GR}. We find that the coordinates choosen by the superspace prescription and those of the bosonic prescription differ and we display the coordinate transformation. We use this example also to illustrate that the dual background may have torsion even if the original one does not.

The action of the $O(3)$ model is
\begin{equation}
S=\int d^{2}x\partial_{\mu}\sigma^{a}\partial^{\mu}\sigma^{a}
\end{equation}
with the constraint that $\sigma^{a}\sigma^{a}=1$. In coordinates $\varphi\equiv(\sigma^{1}+i\sigma^{2})/(1+\sigma^{3})$ that solve the constraint it becomes
\begin{eqnarray}
S&= &\int d^{2}x\left(\frac{1}{1+\varphi\bar{\varphi}}\right)^{2}\partial_{\mu}\varphi\partial^{\mu}\bar{\varphi} \nonumber \\
&=&\int d^{2}xG_{\varphi\bar{\varphi}}\partial_{\mu}\varphi\partial^{\mu}\bar{\varphi}. \label{bosaction}
\end{eqnarray}

This $\sigma$-model has a {\em K\"{a}hler structure}. In fact
\begin{equation}
G_{\varphi\bar{\varphi}}=\partial_{\varphi}\partial_{\bar{\varphi}}\ln(1+\varphi\bar{\varphi})\equiv\partial_{\varphi}\partial_{\bar{\varphi}}K(\varphi ,\bar{\varphi}) \label{Kahler-pot}
\end{equation}
where $K(\varphi ,\bar{\varphi})$ is the K\"{a}hler potential.

An $N=2$ susy version is given by
\begin{equation}
S_{N=2}=\int d^{2}x d^{2}\theta d^{2}\bar{\theta}\ln(1+\phi\bar{\phi})
\end{equation}
with chiral\footnote{See section \ref{susysec}.} superfield $\phi=\varphi_{1}+i\varphi_{2}$, and where $\varphi_{1}$ and $\varphi_{2}$ transform under the $U(1)$-symmetry according to;
\begin{equation}
\begin{split}
\varphi_{1}&\rightarrow\varphi_{1}\cos\theta -\varphi_{2}\sin\theta \\
\varphi_{2}&\rightarrow\varphi_{2}\cos\theta +\varphi_{1}\sin\theta \\
\end{split}
\end{equation}

\subsection{Dualization of the bosonic model}

To familiarize the reader with the dualization procedure in this example, we dualize in two sets of coordinates.

In real coordinates the $O(3)$ model action is written as
\begin{equation}
S=\int d^{2}x\left(\frac{1}{1+\varphi_{1}^{2}+\varphi_{2}^{2}}\right)^{2}\left[(\partial\varphi_{1})^{2}+(\partial\varphi_{2})^{2}\right]. \label{O3-action}
\end{equation}
Using polar coordinates $(\varphi ,\theta)$, (adapted to the $U(1)$ rotation), letting $\partial\theta\rightarrow V$ and $\varphi^2\equiv\varphi_{1}^{2}+\varphi_{2}^{2}$, the parent action may be written as
\begin{equation}
S_{P}=\int d^{2}x\left\{\left(\frac{1}{1+\varphi^{2}}\right)^{2}\left[\varphi^{2}V^2+(\partial\varphi)^{2}\right]+2\lambda\epsilon^{\nu\mu}\partial_{\nu}V_{\mu}\right\}. \label{O3-parent}
\end{equation}
Variation of this action with respect to $\lambda$ yields
\begin{equation}
\delta\lambda :~~~~~V_{\mu}=\partial_{\mu}\theta.
\end{equation}
Putting the expression for $V_{\mu}$ back into the parent action gives the original action $S$ (in adapted coordinates).

Variation with respect to $V_{\mu}$ yields
\begin{equation}
\delta\ V_{\mu}:~~~~~\varphi^{2}GV^{\mu}=\epsilon^{\nu\mu}\partial_{\nu}\lambda~~~\Rightarrow ~~~V^{\mu}=\varphi^{-2}G^{-1}\epsilon^{\nu\mu}\partial_{\nu}\lambda \label{O3-motion}
\end{equation}
where $G\equiv 1/(1+\varphi^2)^{2}$. The action dual to (\ref{O3-action}) that we find by plugging (\ref{O3-motion}) back into (\ref{O3-parent}), is 
\begin{eqnarray}
\tilde{S}&=&\int d^{2}x\left\{G(\partial\varphi)^{2}+\varphi^{-2}G^{-1}(\partial\lambda)^{2}\right\} \nonumber \\
&=&\int d^{2}x\left\{\left(\frac{1}{1+\varphi^{2}}\right)^{2}(\partial\varphi)^{2}+\frac{(1+\varphi^{2})^{2}}{\varphi^{2}}(\partial\lambda)^{2}\right\}
\end{eqnarray}
This form of the dual action is difficult to compare to the (bosonic part of the) dual action in superspace. Let us therefore repeat the above procedure in a different set of coordinates. Defining
\begin{equation}
\varphi\equiv\exp\omega
\end{equation}
the action (\ref{bosaction}) transforms into
\begin{equation}
S=\int d^{2}x\frac{\exp(\omega+\bar{\omega})}{(1+\exp(\omega+\bar{\omega}))^{2}}\partial_{\mu}\omega\partial^{\mu}\bar{\omega} \label{action-new}
\end{equation}
If we further introduce $\rho$ and $\kappa$ via
\begin{equation}\omega\equiv\frac{1}{2}(\rho +i\kappa)~~~\Rightarrow ~~~\omega+\bar{\omega}=\rho ~~;\omega-\bar{\omega}=i\kappa ,
\end{equation}
the action (\ref{action-new}) becomes
\begin{equation}
S=\int d^{2}x\frac{1}{4}\frac{\exp\rho}{(1+\exp\rho)^{2}}\left[(\partial\rho)^{2}+(\partial\kappa)^{2}\right].
\end{equation}
The parent action is now
\begin{equation}
S_{P}=\int d^{2}x\left\{\frac{1}{4}\frac{\exp\rho}{(1+\exp\rho)^{2}}\left[(\partial\rho)^{2}+V^2\right]+\frac{1}{2}\lambda\epsilon^{\nu\mu}\partial_{\nu}V_{\mu}\right\}.
\end{equation}
Varying the parent action with respect to the $V_{\mu}$ yields
\begin{equation}
\delta V_{\mu}:~~~~~\frac{\exp\rho}{(1+\exp\rho)^{2}}V^{\mu}=\epsilon^{\nu\mu}\partial_{\nu}\lambda .
\end{equation}
Putting the corresponding expression for $V_{\mu}$ back into the parent action gives dual action in the new coordinates
\begin{equation}
S_{P}\rightarrow \tilde{S}=\frac{1}{4}\int d^{2}x\left\{\frac{\exp\rho}{(1+\exp\rho)^{2}}(\partial\rho)^{2}+\frac{(1+\exp\rho)^{2}}{\exp\rho}(\partial\lambda)^{2}\right\} \label{parent-bosaction}
\end{equation}

\subsection{The N=2 supersymmetric model} \label{susysec}

Supersymmetric non-linear $\sigma$-models are closely related to complex geometry \cite{kn:Z,kn:AF}. In two dimensions the target space geometry must be K\"{a}hler when $N=2$ and hyperk\"{a}hler when $N=4$. Inclusion of torsion potential terms changes this classification a bit \cite{kn:GHR}, but still restricts the geometry. 

There is no restriction on the target space for the $N=1$ supersymmetric $\sigma$-model. The action is written\footnote{We use coordinates $z,\bar{z}=x^{1}\pm x^{2}$, $d^{2}z\equiv dzd\bar{z}$, $\partial\equiv\partial/\partial z$.}
\begin{equation}
S_{N=1}=\int d^2z d^2\theta(G_{ij}+B_{ij})D_{+}\phi^{i}D_{-}\phi^{j}
\end{equation}
where $\phi^{i}\rightarrow\phi^{i}(z,\bar{z},\theta)$, $D_{+}^{2}=\partial$ and $D_{-}^{2}=\bar{\partial}$.

In the $N=2$ model the target space must be K\"{a}hler. The action is written
\begin{equation}
S_{N=2}=\int d^{2}z d^{2}\theta d^{2}\bar{\theta}K(\phi^{i},\bar{\phi^{i}},\chi^{i},\bar{\chi^{i}})
\end{equation}
where $K$ is the (generalized) K\"{a}hler potential. The superfields satisfy the conditions
\begin{eqnarray}
\bar{D}_{\pm}\phi=D_{\pm}\bar{\phi}&=&0\quad\mbox{(the chiral condition)} \\
\bar{D}_{+}\chi=D_{-}\chi&=&0\quad\mbox{(the twisted chiral condition).}
\end{eqnarray}

The $N=2$ supersymmetry is defined through the (anti) commutation relations $\{D_{+},\bar{D}_{+}\}=\partial$, $\{D_{-},\bar{D}_{-}\}=\bar{\partial}$. The two spinor indices are written out explicitly as $+$ and $-$.

We now return to the $O(3)$ model, with K\"{a}hler potential as in (\ref{Kahler-pot}). In superspace the K\"{a}hler potential is
\begin{equation}
K(\phi+\bar{\phi})=\ln(1+\exp(\phi+\bar{\phi}))
\end{equation}
 and the action is
\begin{equation}
S_{N=2}=\int d^{2}z d^{2}\theta d^{2}\bar{\theta}\ln(1+\exp(\phi+\bar{\phi})) \label{susy-act}.
\end{equation}
The parent action to (\ref{susy-act}) is
\begin{equation}
S_{P}=\int d^{2}z d^{2}\theta d^{2}\bar{\theta}\left\{\ln(1+\exp X)-X(\Lambda+\bar{\Lambda})\right\},
\end{equation}
where $\Lambda$ is a twisted chiral superfield. The equivalence to (\ref{susy-act}) is seen from the field equations for $\Lambda$ and $\overline{\Lambda}$ which say that $X=\phi+\overline{\phi}$ for a chiral field $\phi$. Varying the parent action with respect to $X$ we find
\begin{eqnarray}
\delta X:~~~~~\left(\frac{1}{1+\exp X}\right)\exp X&=&\Lambda+\bar{\Lambda} \nonumber \\
\exp X&=&\frac{\Lambda+\bar{\Lambda}}{1-(\Lambda+\bar{\Lambda})}.
\end{eqnarray}
Putting the expression for $X$ back into the parent action, the action turns into, (we define $\Lambda+\bar{\Lambda}\equiv x$)
\begin{eqnarray}
S_{P}\rightarrow \tilde{S}_{N=2}&=&\int d^{2}z d^{2}\theta d^{2}\bar{\theta}\left\{\ln\left[1+\frac{x}{1-x}\right]-\ln\left(\frac{x}{1-x}\right)x\right\} \nonumber \\
&=&\int d^{2}z d^{2}\theta d^{2}\bar{\theta}\left\{-(1-x)\ln(1-x)-x\ln x\right\}. \label{N=2-action}
\end{eqnarray}
The metric is found by differentiating $K$ twice
\begin{eqnarray}
K&=&-\{x\ln x+(1-x)\ln(1-x)\} \nonumber \\
K_{x}&=&-\{\ln x+1-\ln(1-x)-1\} \nonumber \\
K_{xx}&=&-\left\{\frac{1}{x}+\frac{1}{1-x}\right\}=-\frac{1}{x(1-x)}=g_{xx} \label{K-metric}
\end{eqnarray}

From (\ref{K-metric}) we read off the bosonic part of the action (\ref{N=2-action}) (returning to real coordinates):
\begin{eqnarray}
S&=&\int d^{2}x K_{\varphi\bar{\varphi}}\partial_{a}\varphi\partial^{a}\bar{\varphi} \nonumber \\
&=&\int d^{2}x g_{xx}(\partial_{a}\varphi_{0}\partial^{a}\varphi_{0}+\partial_{a}\varphi_{1}\partial^{a}\varphi_{1}) \label{bosonic-part}
\end{eqnarray}
where $\varphi\equiv\varphi_{0}+i\varphi_{1}$ is the lowest component of the superfield $\phi$.

To compare to our previous result, we return to the bosonic dual action (\ref{parent-bosaction})
\begin{equation*}
S_{P}=\frac{1}{4}\int d^{2}x\left\{\frac{\exp\rho}{(1+\exp\rho)^{2}}(\partial\rho)^{2}+\frac{(1+\exp\rho)^{2}}{\exp\rho}(\partial\lambda)^{2}\right\}
\end{equation*}
We want this on the form (\ref{bosonic-part}). Clearly a change of coordinates (a field redefinition) is needed. We try 
\begin{equation}
\tilde{\rho}=\frac{1}{1+\exp\rho}
\end{equation}
This transforms the action into
\begin{equation}
S_{P}\rightarrow\frac{1}{4}\int d^{2}x\frac{1}{\tilde{\rho}(1-\tilde{\rho})}[(\partial\tilde{\rho})^{2}+(\partial\lambda)^{2}]
\end{equation}
where the metric
\begin{equation}
G=-\frac{1}{\tilde{\rho}(1-\tilde{\rho})}
\end{equation}
So, we recover the action (\ref{bosonic-part}). We see that the superspace version of duality automatically leads to complex coordinates whereas the complex manifold structure gets obscured when we use the Buscher rules.

\subsection{The O(3) model with a $\theta$-term}

In this subsection we include a $B$-term in the $O(3)$ action. We call it a $\theta$-term since in this case it is topological (i.e. a total divergence).

The action for the $O(3)$ $\sigma$-model is
\begin{equation}
S=\frac{1}{g^2}\int d^{2}x\left\{\partial_{a}\sigma^{i}\partial^{a}\sigma^{i}-\alpha(\sigma^{i}\sigma^{i}-1)+g^2\theta\epsilon^{ijk}\epsilon^{ab}\sigma^{i}\partial_{a}\sigma^{j}\partial_{b}\sigma^{k}\right\} \label{O3-model}
\end{equation}
where $i,j \ldots$ denote $O(3)$ indices.
Using stereographic projection
\begin{equation}
\varphi\equiv\frac{\sigma^{1}+i\sigma^{2}}{1+\sigma^{3}}
\end{equation}
we have
\begin{equation}
\sigma^{1}=\frac{\varphi+\bar{\varphi}}{1+\varphi\bar{\varphi}};~~~\sigma^{2}=\frac{i(\bar{\varphi}-\varphi)}{1+\varphi\bar{\varphi}};~~~\sigma^{3}=\frac{1-\varphi\bar{\varphi}}{1+\varphi\bar{\varphi}}
\end{equation}
In these coordinates the action becomes
\begin{equation}
S=\int d^{2}x\left\{\frac{1}{g^2(1+\varphi\bar{\varphi})^{2}}\partial_{a}\bar{\varphi}\partial^{a}\varphi-\frac{4i\theta}{(1+\varphi\bar{\varphi})^{2}}\epsilon^{ab}\partial_{a}\bar{\varphi}\partial_{b}\varphi\right\} \label{newcoaction}
\end{equation} 
Defining $\varphi\equiv\varphi_{0}+i\varphi_{1}$, the Buscher rules give us
\begin{equation}
\begin{split}
\tilde{G}_{00}&=G_{00}^{-1}=g^2(1+\varphi_{1}^{2})^{2} \\
\tilde{G}_{01}&=G_{00}^{-1}B_{01}=4\theta g^2 \\
\tilde{G}_{11}&=G_{11}-G_{00}^{-1}B_{10}B_{01}=\frac{1}{g^2(1+\varphi_{1}^2)^2}\left(1+16\theta^2 g^4\right) \\
\tilde{B}_{01}&=G_{00}^{-1}G_{01}=0 \\
\end{split}
\end{equation}
So, the action (\ref{newcoaction}) dualizes to a model without a $\theta$-term:
\begin{equation}
\tilde{S}=\int d^2{x}\left\{g^2(1+\varphi_{1}^2)^2(\partial\tilde{\varphi_{0}})^2+8\theta g^2\partial_{a}\tilde{\varphi_{0}}\partial^{a}\varphi_{1}+\frac{(1+16\theta^2 g^4)}{g^2(1+\varphi_{1}^{2})^2}(\partial\varphi_{1})^2\right\}
\end{equation}

In the string context this means that a torsionful string background may dualize to one without torsion, thus changing the geometry drastically. This may not be totally obvious from the present simple example, since the torsion term in (\ref{O3-model}) is a total divergence, and thus not really there (in trivial topologies). A more interesting and nontrivial example of $N=2$ duality is given by the $WZW$ model on $SU(2)\times U(1)$ which is dual to $[SU(2)/U(1)]\times U(1)^{2}$ \cite{kn:A,kn:B}.

\section{Non-Abelian dualization}

At the end of our disussion of $S$-duality we gave a $3D$ example that has a non-Abelian generalization. We mentioned then that this is a fairly rare non-Abelian case. When it comes to $S$-duality, however, there are general prescriptions. We summarize one of them \cite{kn:OQ} below.

We start from the action
\begin{equation}
S=\int d^{2}x G_{\hat{i}\hat{j}}(\phi)\partial_{a}\phi^{\hat{i}}\partial^{a}\phi^{\hat{j}},~~~~~\hat{i}=(i,\mu) \label{non-A-action}
\end{equation}
which is assumed to have an symmetry group $G$ of non-Abelian isometries
\begin{equation}
\phi^{i}\rightarrow g^{i}_{~j}\phi^{j};~~~~~g\in\mathcal{G}
\end{equation}
We gauge a subgroup $H\subseteq\mathcal{G}$ introducing the corresponding gauge potential $A_{\mu}$. The parent action may be written 
\begin{equation}
S_{P}=\int d^{2}x\left\{G_{ij}(\phi)D_{a}\phi^{i}D^{a}\phi^{j}+\mbox{tr}\Lambda F\right\} \label{non-A-P-action}
\end{equation}
where the covariant derivative is 
\begin{equation}
D_{a}\phi^{i}=\partial_{a}\phi^{i}+A^{\alpha}_{a}(T_{\alpha})^{i}_{j}\phi^{j};~~~~~A_{a}\rightarrow(A,\bar{A})
\end{equation}
with $T_{\alpha}$ the generators of the Lie algebra of $G$.
The field strength is 
\begin{equation}
F\equiv\partial\bar{A}-\bar{\partial}A+[A,\bar{A}].
\end{equation}
Variation of the parent action with respect to the Lagrange multiplier $\Lambda$ yields
\begin{equation}
\delta\Lambda :~~~~~F=0
\end{equation}
so that the gauge potential is ``pure gauge''
\begin{equation}
A=h^{-1}\partial h,~~~~~\bar{A}=h^{-1}\bar{\partial}h,~~~~~h\in H
\end{equation}
Plugging this back into (\ref{non-A-P-action}) we recover (\ref{non-A-action}).
The dual action found by eliminating $A_{\mu}$ is
\begin{equation}
\tilde{S}_{\phi,\Lambda}=S_{\phi}-\int d^{2}z\bar{h}_{\alpha}(f^{-1})^{\alpha\beta}h_{\beta}
\end{equation}
where
\begin{equation}
\begin{split}
f_{\alpha\beta}&\equiv -c_{\alpha\beta}^{~~\gamma}\Lambda_{\gamma}+\phi^{i}(T_{\beta})^{k}_{~i}G_{kl}(T_{\alpha})^{l}_{~m}\phi^{m} \\ \\
h_{\alpha}&\equiv-\partial\Lambda_{\alpha}+G_{ai}\partial\phi^{a}(T_{\alpha})^{i}_{~v}\partial\phi^{i}
\end{split}
\end{equation}
But taking the dual of the dual does not give the original action back\footnote{$D^2\neq 1$ (cf. excercise at the end of chapter 1).}. In fact the dual metric may not have any isometries at all. In the next section, however, we will show how it may be possible to solve this problem.

\section{Poisson-Lie T-duality}

In this last section we will introduce the basics of what has been called {\it Poisson-Lie T-duality} \cite{kn:KS}. It generalizes the abelian and the traditional non-abelian dualities (the latter was presented in the previous section) since this construction is not based on the presence of isometries for the background and the dual background metrics.

Consider a $\sigma$-model with fields $\phi^{i}$ mapping fields $\phi^{i}$ from some two dimensional worldsheet manifold $\mathcal{M}$ into a target space T. The target space is assumed to have a metric $G_{ij}$ as well as a torsion potential $B_{ij}$. The action may be written
\begin{equation}
S_{\phi}=\int d^2 z\partial \phi^{i}F_{ij}(\phi)\bar{\partial}\phi^{j} \label{PL-action}
\end{equation}
where $F_{ij}=G_{ij}+B_{ij}$. The group structure $G$ of the target space $T$ defines a group action
\begin{equation}
\delta \phi^{i}=k^{a}e^{i}_{a} \label{group-action}
\end{equation}
where $k^{a}$ are parameters that depend on the coordinates ($z,\bar{z}$) of $\mathcal{M}$ and $e^{i}_{a}$ are the invariant frame fields\footnote{In three and four dimensions the frame fields are usually called triads (dreibeins) or tetrads (vierbeins), respectively.}  in the Lie algebra $\mathcal{G}$ of the group $G$; $a=1,2,\ldots ,\mbox{dim}G$. The frame fields obey the relation
\begin{equation}
[e_{a},e_{b}]^{i}=f^{c}_{ab}e^{i}_{c}
\end{equation}
where $f^{c}_{ab}$ are the structure constants of the Lie group $G$.

Varying the action with respect to $\phi^{i}$ gives
\begin{equation}
\delta S_{\phi}=\int d^{2}z\{k^{a}\partial\phi^{i}\pounds_{e_{a}}F_{ij}\bar{\partial}\phi^{j}+\partial k^{a}e^{i}_{a}F_{ij}\bar{\partial}\phi^{j}+\bar{\partial}k^{a}\partial\phi^{i}F_{ij}e^{j}_{a}\} \label{action-var}
\end{equation}
where $\pounds_{k_{a}}F_{ij}$ is the Lie derivatives of $F_{ij}$. The currents $J_{a}$ are given by the 1-form
\begin{equation}
{\cal J}_{a}=J_{a}dz+\bar{J}_{a}d\bar{z}
\end{equation}
where\footnote{We may find the currents from the Lagrangian using the definition $J^{\mu}=(\partial{\cal L}/\partial(\partial_{\mu}\phi^{i}))\delta\phi^{i}$, where $\mu=\{z,\bar{z}\}$. Remembering that $J^{z}=J_{\bar{z}}\equiv\bar{J}$ and $J^{\bar{z}}=J_{z}\equiv J$, we find the results given in eq.(\ref{currents}).}
\begin{equation}
J_{a}=\partial \phi^{i}F_{ij}e^{j}_{a}~~;~~~~~\bar{J}_{a}=e^{i}_{a}F_{ij}\bar{\partial}\phi^{j} \label{currents}
\end{equation}

From eq.(\ref{action-var}) we thus see that a target space with isometries, i.e. $\pounds_{e_{a}}F_{ij}=0$\footnote{In section \ref{isometries} it was shown that if the metric has an isometry, that isometry is a symmetry of the $\sigma$-model action.}, gives us simply the equations of motion
\begin{equation} 
\partial\bar{J}_{a}+\bar{\partial}J_{a}=0
\end{equation}
Written in a coordinate independent way we have $d\ast {\cal J}_{a}=0$, where $\{\ast {\cal J}_{a}\}$ is the dual basis to $\{{\cal J}_{a}\}$. This is a special case corresponding to $\tilde{f}^{c}_{ab}=0$, where $\tilde{f}^{c}_{ab}$ are the structure constants of the dual target space (with Lie algebra $\tilde{\mathcal{G}}$). However, in the non-Abelian case, the equations of motion are given by the Maurer-Cartan equation
\begin{equation}
d\ast {\cal J}_{a}+\frac{1}{2}\tilde{f}_{a}^{bc}\ast {\cal J}_{b}\wedge\ast {\cal J}_{c}=0 \label{Maur-Cart}
\end{equation}
The component equations of (\ref{Maur-Cart}) read
\begin{equation}
\partial\bar{J}_{a}+\bar{\partial}J_{a}+J_{b}\tilde{f}_{a}^{bc}\bar{J}_{c}=0 \label{comp-eq}
\end{equation}
From (\ref{action-var}) we find $\partial\phi^{i}\pounds_{e_{a}}F_{ij}\bar{\partial}\phi^{j}=\partial\bar{J}_{a}+\bar{\partial}J_{a}$. If we further use the expression for the currents (\ref{currents}) the component equations (\ref{comp-eq}) give the following condition for $F_{ij}$:
\begin{equation}
\pounds_{e_{a}}F_{ij}=F_{ik}e^{k}_{b}\tilde{f}^{bc}_{a}e^{l}_{c}F_{lj} \label{cond-metric}
\end{equation}

We know that the Lie derivatives have to satisfy the relation
\begin{equation}
[\pounds_{e_{a}},\pounds_{e_{b}}]=f_{ab}^{c}\pounds_{e_{c}} \label{comm-rel}
\end{equation}
Applying eq.(\ref{comm-rel}) to (\ref{cond-metric}) thus gives us the relation between the structure constants of the target space and the structure constants of the dual target space
\begin{equation}
f^{a}_{dc}\tilde{f}^{rs}_{a}=\tilde{f}^{as}_{c}f^{r}_{da}+\tilde{f}^{ra}_{c}f^{s}_{da}-\tilde{f}^{as}_{d}f^{r}_{ca}-\tilde{f}^{ra}_{d}f^{s}_{ca}
\end{equation}
This is the Jacobi identity for the Lie bi-algebra $({\cal G},{\cal \tilde{G}})$, the so-called {\it Drinfeld Double}.

We now expect that the action dual to (\ref{PL-action}),
\begin{equation}
\tilde{S}_{\tilde{\phi}}=\int d^{2}z\partial \tilde{\phi}^{i}\tilde{F}_{ij}(\tilde{\phi})\bar{\partial}\tilde{\phi}^{j},
\end{equation}
should obey the same condition as (\ref{cond-metric}) but with the tilded and un-tilded variables interchanged
\begin{equation}
\pounds_{\tilde{e}^{a}}\tilde{F}^{ij}=\tilde{F}^{ik}\tilde{e}^{b}_{k}f^{a}_{bc}\tilde{e}^{c}_{l}\tilde{F}^{lj} \label{dual-PLcond}
\end{equation}
The backgrounds we expect to be related via
\begin{equation}
(F(\phi =0))_{ij}^{-1}=\tilde{F}^{ij}(\tilde{\phi}=0) \label{extra-cond}
\end{equation}

Now, we want the field equations (\ref{Maur-Cart}) of the original theory to be the Bianchi identity of the dual theory. This is achieved as follows. We introduce the Maurer-Cartan form $\ast {\cal J}$ on $\tilde{G}$ which we expand in terms of the dual basis $\{ \tilde{T}^{a}\}$
\begin{equation}
\ast {\cal J}=\ast {\cal J}_{a}\tilde{T}^{a}
\end{equation}
where $[\tilde{T}^{b},\tilde{T}^{c}]=\tilde{f}^{bc}_{a}\tilde{T}^{a}$. If $\ast {\cal J}$ is a pure gauge field, i.e.
\begin{equation}
\ast {\cal J}\equiv\tilde{g}^{-1}d\tilde{g};~~~~~\tilde{g}\in\tilde{\cal G},
\end{equation}
this ensures that the field equations (\ref{Maur-Cart}) becomes the Bianchi identity in the dual theory.

To summarize, we have learned that two target spaces are dualizeable only if the backgrounds satisfy the system of partial differential equations given in (\ref{Maur-Cart}) and (\ref{dual-PLcond}). In addition the backgrounds must satisfy the condition (\ref{extra-cond}). In other words the original theory and the ``dual'' theory can be dual to each other if their target spaces can be embedded into a Drinfeld Double.

A general feature is that backgrounds without torsion are related to backgrounds with torsion. Furthermore, from eq.(\ref{Maur-Cart}) and (\ref{dual-PLcond}) we see that an abelian theory ($f^{a}_{bc}=0$) gives a dual theory of non-abelian isometries ($\pounds_{e_{a}}\tilde{F}_{ij}=0$), and vice versa. In a quantum analysis of the Poisson-Lie T-duality, it appears that when $G$ and $\tilde{G}$ are both non-abelian, one has to shift the dilaton in both theories to maintain conformal invariance. 

\chapter*{~Further reading}
\addcontentsline{toc}{chapter}{Further reading}

In these lectures we have presented the basics of duality. The role of the duality in modern theoretical physics seems to be a very important one at the moment. A natural continuation for the reader who is interested in the subject is to learn about the generalized EM duality of the Montonen-Olive type in supersymmetric gauge theories. There are many good reviews of this subject and we recommend the ones by Olive \cite{kn:OLI}, G\'{o}mez and Hern\'{a}ndez \cite{kn:GOM}, Harvey \cite{kn:HARV} and Di Vecchia \cite{kn:DIV}. We also recommend four techniqual papers on duality: \cite{kn:GR1}, \cite{kn:GR2}, \cite{kn:RV} and \cite{kn:DGHT}. 

\vspace*{3cm}

\begin{flushleft}
{\large\bf{~Acknowledgements}}
\addcontentsline{toc}{chapter}{Acknowledgements}
\end{flushleft}
\noindent
We would like to thank M. Ro\v{c}ek and R.v.Unge for reading and commenting on this manuscript. U.L. also acknowledges partial support from NFR under grant no. 4038-312 and from NorFA under grant no. 96.55.030-O.


\appendix

\chapter{Notation and Conventions}

We use the Minkowski metric $g_{\mu\nu}=\mbox{diag}(+1,-1,\ldots ,-1)$.

In $D=4$ the Levi-Civita symbol $\epsilon^{\mu\nu\rho\sigma}$ (with the definition $\epsilon^{0123}=1$) obey the following relations:
\begin{equation}
\begin{split}
\epsilon^{\mu\nu\rho\sigma}\epsilon_{\mu\nu\rho\sigma}&=-4!   \\
\epsilon^{\mu\nu\rho\sigma}\epsilon_{\lambda\nu\rho\sigma}&=-3!\delta^{\mu}_{~\lambda} \\
\epsilon^{\mu\nu\rho\sigma}\epsilon_{\lambda\kappa\rho\sigma}&=-2!\delta^{\mu}_{~[\lambda}\delta^{\nu}_{~\kappa]} \\
\epsilon^{\mu\nu\rho\sigma}\epsilon_{\lambda\kappa\tau\sigma}&=-1!\delta^{\mu}_{~[\lambda}\delta^{\nu}_{~\kappa}\delta^{\rho}_{~\tau]}  \\  
\end{split}
\end{equation}

In $D=3$ and $D=2$ corresponding relations for the Levi-Civita symbol (with the definitions $\epsilon^{012}=1$ and $\epsilon^{01}=1$, respectively) can be found from the relations
\begin{equation}
\epsilon^{\mu\nu\rho}\epsilon_{\lambda\kappa\tau}=\delta^{\mu}_{[\lambda}\delta^{\nu}_{~\kappa}\delta^{\rho}_{\tau]},~~~~~\epsilon^{\mu\nu}\epsilon_{\lambda\kappa}=-\delta^{\mu}_{[\lambda}\delta^{\nu}_{\kappa]}
\end{equation}

The Hodge dual is defined as follows:
\begin{equation}
\begin{split}
\ast F^{\mu\nu\rho}&\equiv \epsilon^{\mu\nu\rho\sigma}F_{\sigma} \\
\ast F^{\mu\nu}&\equiv \frac{1}{2!}\epsilon^{\mu\nu\rho\sigma}F_{\rho\sigma} \\
\ast F^{\mu}&\equiv \frac{1}{3!}\epsilon^{\mu\nu\rho\sigma}F_{\nu\rho\sigma}
\end{split}
\end{equation}

The anti-symmetrization rule is defined as
\begin{equation}
\begin{split}
A_{[\mu}B_{\nu]}&\equiv A_{\mu}B_{\nu}-A_{\nu}B_{\mu} \\
A_{[\mu}B_{\nu}C_{\rho]}&\equiv A_{\mu}B_{[\nu}C_{\rho]}-A_{\nu}B_{[\mu}C_{\rho]}+A_{\rho}B_{[\mu}C_{\nu]} \\
&=A_{\mu}B_{\nu}C_{\rho}-A_{\mu}B_{\rho}C_{\nu}-A_{\nu}B_{\mu}C_{\rho} \\
&+A_{\nu}B_{\rho}C_{\mu}+A_{\rho}B_{\mu}C_{\nu}-A_{\rho}B_{\nu}C_{\mu} \\
\end{split}
\end{equation}

Some useful relations are
\begin{equation}
\begin{split}
\epsilon^{\alpha\mu\nu\rho}\partial_{[\mu}A_{\nu\rho]}&=3!\epsilon^{\alpha\mu\nu\rho}\partial_{\mu}A_{\nu\rho} \\
\delta^{\mu}_{~[\lambda}\delta^{\nu}_{~\kappa}\delta^{\rho}_{~\tau]}\partial_{\mu}A_{\nu\rho}&=\partial_{[\lambda}A_{\kappa\tau]} \\
\delta^{\rho}_{~[\lambda}\delta^{\mu}_{~\kappa}\delta^{\nu}_{~\alpha]}\partial^{\lambda}A^{\kappa\alpha}&=3!\partial^{[\rho}A^{\mu\nu]} \\
\end{split}
\end{equation}

\clearemptydoublepage


\bibliographystyle{unsrt}

\end{document}